\documentclass{article}
\pdfoutput=1

\usepackage{arxiv}

\usepackage[utf8]{inputenc} 
\usepackage[T1]{fontenc}    
\usepackage{hyperref}       
\usepackage{url}            
\usepackage{booktabs}       
\usepackage{amsfonts}       
\usepackage{nicefrac}       
\usepackage{microtype}      
\usepackage{lipsum}
\usepackage{amsmath}
\DeclareMathOperator*{\argminB}{argmin}
\usepackage{mathtools}

\usepackage{multirow}
\usepackage[round]{natbib}
\usepackage{placeins}

\usepackage[english]{babel}
\usepackage{amsthm}
\theoremstyle{definition}
\newtheorem{definition}{Definition}[section]

\newcommand\independent{\protect\mathpalette{\protect\independenT}{\perp}}
\def\independenT#1#2{\mathrel{\rlap{$#1#2$}\mkern2mu{#1#2}}}

\newtheorem{algorithm}{Algorithm}[section]

\title{Knockoff Boosted Tree for Model-Free Variable Selection}

\author{
  Tao Jiang\\
  North Carolina State University\\
  \texttt{tjiang8@ncsu.edu} \\
  \And
  Yuanyuan Li\\
  National Institute of Environmental Health Sciences\\
  \And
  Alison A. Motsinger-Reif\thanks{To whom correspondence should be addressed.}\\
  National Institute of Environmental Health Sciences\\
  \texttt{motsingerreifaa@nih.gov} \\
}

\begin{document}
\maketitle

\begin{abstract}
In this article, we propose a novel strategy for conducting variable selection without prior model topology knowledge using the knockoff method with boosted tree models. Our method is inspired by the original knockoff method, where the differences between original and knockoff variables are used for variable selection with false discovery rate control. The original method uses Lasso for regression models and assumes there are more samples than variables. We extend this method to both model-free and high-dimensional variable selection. We propose two new sampling methods for generating knockoffs, namely the sparse covariance and principal component knockoff methods. We test these methods and compare them with the original knockoff method in terms of their ability to control type I errors and power. The boosted tree model is a complex system and has more hyperparameters than models with simpler assumptions. In our framework, these hyperparameters are either tuned through Bayesian optimization or fixed at multiple levels for trend detection. In simulation tests, we also compare the properties and performance of importance test statistics of tree models. The results include combinations of different knockoffs and importance test statistics. We also consider scenarios that include main-effect, interaction, exponential, and second-order models while assuming the true model structures are unknown. We apply our algorithm for tumor purity estimation and tumor classification using the Cancer Genome Atlas (TCGA) gene expression data. The proposed algorithm is included in the KOBT package, available at \url{https://cran.r-project.org/web/packages/KOBT/index.html}.
\end{abstract}

\keywords{Boosted tree model \and Knockoff \and Shapley additive explanation \and Model-free variable selection}

\section{Introduction}

In nearly all existing variable selection methods, a set of predictor candidates must be specified before selection \citep{li2005model}. Predictors in the final selected model are a subset of candidate set. However, it can be challenging to determine an appropriate set of predictor candidates, especially considering the order (e.g., quadratic, cubic) of main effects and interactions with a large number of variables. \cite{li2005model} stated the difference between model selection and variable selection is whether to assume a model for comparison. One of the advantages of model-free variable selection is that its performance is not limited by the choice of model formulation. Model-free methods such as Random Forest \citep{breiman2001random}, AdaBoost \citep{ratsch2001soft}, and convolutional neural networks \citep{lecun1995convolutional} are widely used in research and industry for regression and classification purposes. These methods recognize data patterns without the need to impose specific model structures on regression functions. This independence allows model-free methods more flexibility, thus making them highly appropriate in many cases. For example, \cite{gao2018model} reported that compared with other methods, model-free methods provide more reliable clinical outcomes when forecasting falls of Parkinson’s patients.

Among the widely used model-free methods, tree-based models have demonstrated greater effectiveness and interpretability than most other learning algorithms. In a standard regression model, an interaction term is not estimated if it is not specified in the model. In contrast, tree-based methods automatically include interactions during tree growing, without the requirement of an \textit{a prioiri} formula preconception \citep{su2011interaction}. Further, tree-based models are more naturally associated with nonlinear interactions \citep{schiltz2018using}, and the branches in a tree model are original interaction variables. \cite{friedman2001greedy} stated that the number of terminal nodes of a boosted tree model should depend on the highest order of the dominant interactions among the variables. This implies that (1) terminal (leaf) nodes in tree models cannot be interpreted simply as main effects of original variables and (2) growth of a tree model is selective.  

Variable selection in tree models is affected by the number of categories of features \citep{kim2001classification}, with variables with more categories preferred. To eliminate bias in variable selection, \cite{loh2002regression} proposed the GUIDE algorithm, which is used to conduct chi-square analysis of residuals and bootstrap calibration of significance probabilities. Gini importance in random forest methods \citep{breiman2001random} is used to measure improvement in the splitting criterion produced by each corresponding variable. To generate an unbiased importance index, \cite{breiman2008random} proposed permutation importance by statistical permutation. In addition to regression tree and random forest models, attention has also been paid to variable selection in boosted tree models, given their superior performance in data pattern recognition. \cite{miller2016finding} extended gradient boosted tree to multivariate tree boosting and performed nonparametric regression to identify important variables. Gain, weight, and cover are three common importance ranking measures in the highly successful XGBoost \citep{chen2016xgboost}. SHapley Additive exPlanation (SHAP) \citep{lundberg2017unified} has also been proposed to interpret model predictions, and its performance has been highly consistent when applied in tree models \citep{lundberg2018consistent}.

While most data scientists are interested in model prediction accuracy, inference from analysis is also important. Originally described for Random Forest \citep{breiman2001random}, the permutation method has been widely used to generate a null distribution of test statistics to detect significance in tree models. Unfortunately, some statistical inference, such as accurate false discovery rate (FDR) control, is not possible with permutation-based methods. For example, in a linear regression problem, a permutation-based construction may underestimate the false discovery proportion (FDP) in cases where the design matrix displays nonvanishing correlations \citep{barber2015controlling}. Under a global null, \textit{p}-values from permutation tests are exact. However, under an alternative hypothesis, rejecting the null based on permutation tests does not necessarily imply a valid rejection \citep{chung2013exact}. To control the FDR in variable selection, \cite{barber2015controlling} proposed a novel statistical framework, the \textit{knockoff} framework, for low-dimensional $(n \geq p)$ fixed design. The \textit{knockoff} framework was later extended to high-dimensional $(n<p)$ and random design scenarios \citep{candes2018panning}. The method generates knockoff variables without the need to collect new data or consider response variables while retaining the correlation structure of the original variables. These knockoff variables are used as a negative control group, and a variable selection procedure is applied for both the original and knockoff variables. Test statistics measuring the importance of both kinds of variables can be compared to compute symmetrized knockoff statistics. Finally, a threshold for these statistics is determined according to the (desired) target FDR level. The goal of using the \textit{knockoff} method is discovering relevant variables while controlling the FDR.

In this article, we propose the knockoff boosted tree (KOBT) algorithm, a model-free variable selection algorithm that includes boosted tree \citep{friedman2001greedy, chen2016xgboost} and model-X knockoffs \citep{candes2018panning}. Model-X knockoff variables are generated with the same correlation structure as the original variables. Test statistics, such as tree SHAP \citep{lundberg2018consistent}, Cover, Frequency, Gain, and Saabas are generated for each pair of original and knockoff variables, and the consistency and accuracy of the produced test statistics are compared. The FDR is controlled using the difference between original and knockoff test statistics. Further, regularization parameters in boosted tree models are tuned through Bayesian optimization \citep{snoek2012practical}.

In the following sections, we describe our analytical framework, including (1) the generation of knockoff variables; (2) fitting and Bayesian optimization in boosted tree regression; and (3) consistency, accuracy, and FDR control in model-free variable selection. We validate our approach with simulation experiments. Finally, we demonstrate our approach on tumor purity estimation and tumor classification using the Cancer Genome Atlas (TCGA) gene expression data.

\section{Methods and Materials}

\subsection{Parametric Functions of a Single Regression Tree}
\label{sec:one-tree}

Denote $\mathbf{y}_{n \times 1}$ as $n$ responses and $\mathbf{X}_{n \times p}$ as $p$ independent variables each with $n$ values. In regression trees, $\mathbf{y}_{n \times 1}$ can be continuous or ordered discrete. The goal is to minimize an objective function, which may contain a loss function component and a regularization component, for a combination of all leaves (terminal nodes) on a tree. Given a tree with $m$ terminal nodes (i.e., the whole sample contains $m$ partitions), if we model the response as a constant in each region, the estimated regression tree function is defined as
\begin{equation}
    \hat{y}_{i} = \hat{f}(\mathbf{x}_{i}) = \sum_{k=1}^m \mathbf{I} \left\{ \mathbf{x}_{i} \in \mathcal{R}_k^p \right\} \mathbf{\beta}_{k},
\end{equation}
where $\mathbf{I}(\cdot)$ is a function of $\mathbf{x}$, and  $\mathbf{I} \left\{ \mathbf{x}_{i} \in \mathcal{R}_k^p \right\} =1$, if the $i$-th row vector of $\mathbf{X}$, $\mathbf{x}_{i} \in \mathcal{R}_k^p$, the $k$-th disjoint partition region and otherwise $\mathbf{I} \left\{ \mathbf{x}_{i} \in \mathcal{R}_k^p \right\}=0$. $\mathcal{R}_k^p$ is a split region or a so-called leaf, $k=1,...,m$. With the squared $L2$ norm loss function, we have estimator
\begin{equation}
    \hat{\mathbf{\beta}} =  \argminB_{\mathbf{\beta}} \sum_{i=1}^n \left\{ y_i - \sum_{k=1}^m \mathbf{I} \left\{ \mathbf{x}_{i} \in \mathcal{R}_k^p \right\} \mathbf{\beta}_{k} \right\}^2,
\end{equation}
and
\begin{equation} \label{1st.dir}
    \partial \sum_{i=1}^n \left\{ y_i - \sum_{k=1}^m \mathbf{I} \left\{ \mathbf{x}_{i} \in \mathcal{R}_k^p \right\} \mathbf{\beta}_{k} \right\}^2 / \partial \beta_k = 0.
\end{equation}
Solving Equation \ref{1st.dir}, the $k$-th estimated parameter of $\mathbf{\beta}$ is
\begin{equation} \label{estimator}
    \hat{\beta}_k = 
    \frac{\sum_{i=1}^n y_i \mathbf{I}_k \left\{ \mathbf{x}_{i} \in \mathcal{R}_k^p \right\}}{\sum_{i=1}^n \mathbf{I}_k \left\{ \mathbf{x}_{i} \in \mathcal{R}_k^p \right\}},
\end{equation}
where $\mathbf{I}_k \left\{ \mathbf{x}_{i} \in \mathcal{R}_k^p \right\} =0$ for a fixed $\mathcal{R}_k^p$. This is the sample mean of $\mathbf{y}$ on that leaf of the regression tree. Meanwhile, to grow a tree model, the data need to be partitioned or split. For the $j$-th variable $\mathbf{x}_j$, we define a split on $\mathbf{x}_j$ with some value $c$ as
\begin{equation}
    \begin{split}
        \mathcal{R}_{k,<}^p &= \{ i: x_{ij} < c | \mathbf{x}_{i} \in \mathcal{R}_k^p \}, \\
        \mathcal{R}_{k,>}^p &= \{ i: x_{ij} \geq c | \mathbf{x}_{i} \in \mathcal{R}_k^p \}.
    \end{split}
\end{equation}
Based on Equation \ref{estimator}, we have two estimates
\begin{equation}
    \begin{split}
        \hat{\beta}_{k,<} = \frac{\sum_{i=1}^n y_i \mathbf{I}_k \left\{ \mathbf{x}_{i} \in \mathcal{R}_{k,<}^p \right\}}{\sum_{i=1}^n \mathbf{I}_k \left\{ \mathbf{x}_{i} \in \mathcal{R}_{k,<}^p \right\}}, \\
        \hat{\beta}_{k,>} = \frac{\sum_{i=1}^n y_i \mathbf{I}_k \left\{ \mathbf{x}_{i} \in \mathcal{R}_{k,>}^p \right\}}{\sum_{i=1}^n \mathbf{I}_k \left\{ \mathbf{x}_{i} \in \mathcal{R}_{k,>}^p \right\}}.
    \end{split}
\end{equation}
Thus, an optimal split is defined as
\begin{equation}
    \hat{c} =  \argminB_{c} \left\{ \sum_{i=1}^n \left( y_{i \in \mathcal{R}_{k,<}^p} - \hat{\beta}_{k,<} \right)^2 + \sum_{i=1}^n \left( y_{i \in \mathcal{R}_{k,>}^p} - \hat{\beta}_{k,>} \right)^2 \right\}.
\end{equation}
To avoid overfitting, similar to penalized regression, a penalty term is added to the cost-complexity criterion of a regression tree model
\begin{equation} \label{l0_norm}
    \sum_{k=1}^m || \mathbf{y}_{i \in \mathbf{I}_{k}} - \mathbf{1} \hat{\beta}_{k} ||_2^2 + \gamma |T|,
\end{equation}
where $T(\mathbf{X}_{n \times p}; \mathbf{\Theta}) = f(\mathbf{X}_{n \times p})$ is a regression tree model, $\mathbf{\Theta}=\{ \mathbf{\mathcal{R}}, \mathbf{\beta}\}$ contains tree structure parameters, $|T|$ is the cardinality of terminal nodes in $T$, and $\gamma \geq 0$ is a tuning parameter.

\subsection{Boosted Tree Methods and Regularization}

To overcome the bias and high-variance problem in a single regression tree, ensemble tree models, such as bagging or boosting structures, have been proposed. Defining a boosted tree model as a sum of $B$ trees in Section \ref{sec:one-tree},
\begin{equation} \label{sum_functions}
    f^{B}(\mathbf{X}) = \sum_{b=1}^B T(\mathbf{X}_{n \times p}; \mathbf{\Theta}_b) = \sum_{b=1}^B f_b(\mathbf{X}_{n \times p}).
\end{equation}
In a forward stagewise algorithm \citep{hastie2005elements}, the $B$-th tree structure is found by solving
\begin{equation}\label{theta_hat}
    \hat{\Theta}_B = \argminB_{\Theta_B} L(\mathbf{y}, f^{B-1}(\mathbf{X})+T(\mathbf{X}, \Theta_B)),
\end{equation}
where $L(\Theta_B; \mathbf{y}, f^{B-1})$ is a loss function based on the $B$-th tree structure, given response $\mathbf{y}$ and previously fitted $B-1$ tree models $f^{B-1}$.

In boosted tree model fitting, the objective function usually contains a loss function and penalty function for tree complexity (i.e., regularization for tree models). The complexity depends on the number of trees in a sequence, the depth of each tree, and the number of terminal nodes in each tree. One way to control the complexity of tree model fitting is to set a minimum number of instances in a single terminal node. For example, the \textit{min\_child\_weight} parameter in XGBoost \citep{chen2016xgboost} controls the minimum sum of instance weight (Hessian) in a terminal node, which is equivalent to a minimum number of instances if all instances have a weight of $1$. For simplicity, here, we assume all weights are $1$. Combining Equation \ref{l0_norm} and Equation \ref{theta_hat}, the $B$-th objective function is defined as
\begin{equation}
    \begin{split}
        Obj(f_B; \mathbf{y}, f^{B-1}) & = L(\mathbf{y}, f^{B-1}(\mathbf{X})+T(\mathbf{X}, \Theta_B)) + \sum_{b=1}^B \Omega(f_b) \\
        & \propto L(\mathbf{y}, f^{B-1}(\mathbf{X})+T(\mathbf{X}, \Theta_B)) + \Omega(f_B) \\
        & = L(\mathbf{y}, f^{B-1}(\mathbf{X})+T(\mathbf{X}, \Theta_B)) + \gamma |T(\mathbf{X}, \Theta_B))|,
    \end{split}
\end{equation}
where $|T(\mathbf{X}, \Theta_B))|$ is the cardinality of terminal nodes in the $B$-th tree, and $\gamma \geq 0$ is a tuning parameter. Similar to elastic net \citep{zou2005regularization}, XGBoost introduces $L1$ and $L2$-norm penalty terms into objective functions. Therefore, the objective function for the $B$-th tree can be updated as
\begin{equation} \label{norm012}
    Obj(f_B; \mathbf{y}, f^{B-1}) \propto L(\mathbf{y}, f^{B-1}(\mathbf{X})+T(\mathbf{X}, \Theta_B)) + \gamma |T| + \lambda || \mathbf{\beta}_B ||_2^2 + \alpha|| \mathbf{\beta}_B ||_1,
\end{equation}
where $\mathbf{\beta}_B$ is a vector of leaf weights in the $B$-th tree, and $\gamma \geq 0$, $\lambda \geq 0$, and $\alpha \geq 0$ are tuning parameters.

Besides setting the minimum number of instances in terminal nodes and applying penalization on node weights, regularization can be achieved by limiting the maximum depth of a tree, applying the maximum number of trees in a boosting sequence, or defining an early stopping criteria for boosting \citep{zhang2005boosting}. Given the numerous parameters and hyperparameters of regularization and tree structure, it is tedious to find the optimal group of parameters by grid search. The parameters and their optimization are summarized in the following section.

\subsection{Regularization Parameters and Bayesian Optimization}

As reviewed by \cite{nielsen2016tree}, there are three kinds of regularization parameters: boosting parameters, randomization parameters, and tree parameters. Boosting parameters include the number of trees $B$ (i.e., the number of boosting iterations) and the learning rate $\eta$. A small step-size of the learning rate has been found to play an important role in the convergence of boosting procedures \citep{zhang2005boosting}. In simulation studies, decreasing the learning rate increased the performance of boosted models \citep{friedman2000additive}. Using a smaller learning rate and thus a relatively larger number of trees has been suggested, given the relationship between the learning rate and the number of boosted models. A drawback of this procedure is increased running time for model fitting. The randomization parameters refer to the row subsampling parameter and the column subsampling parameter. In the random forest algorithm \citep{breiman2001random}, only a random subsample of the training data set is used for building model. \cite{friedman2002stochastic} built on this idea and introduced stochastic gradient boosting for the boosting tree algorithm. Basically, the row subsampling parameter controls the ratio for a subset of the data, and the column subsampling parameter determines the ratio for a subgroup of features in each tree fitting. Both boosting parameters and randomization parameters are used for general regularization since they are used in the optimization of all kinds of boosting models. However, tree parameters are for tree models only. Therefore, in this article, the discussion is focused on tree parameters and their optimization. 

Note that all the tree parameters implicate a tradeoff between bias and variance. In each individual tree, besides the penalization parameters $\gamma$, $\lambda$, and $\alpha$ in Equation \ref{norm012}, the tree parameters also include structure parameters, the maximum depth of the tree, and the minimum sum of observation weights required for each leaf. Given the maximum depth of a tree, $D$, using the decomposition of target function introduced in \cite{friedman1991multivariate}, the $b$-th individual tree function in Equation \ref{sum_functions} can be defined as
\begin{equation} \label{decomposition}
    f_b(\mathbf{X}_{n \times p}) = \sum_{j} g^{(1)}(\mathbf{x}_j) + \sum_{j, k} g^{(2)}(\mathbf{x}_j, \mathbf{x}_k) + \sum_{j, k, l} g^{(3)}(\mathbf{x}_j, \mathbf{x}_k, \mathbf{x}_l) + ... + \sum_{j, k, l, ...} g^{(D)}(\mathbf{x}_j, \mathbf{x}_k, \mathbf{x}_l, ...),
\end{equation}
where $g^{(1)}(\mathbf{x}_j)$ is the first-order interaction (main effect) of the $j$-th variable, $\mathbf{x}_j$, $g^{(2)}(\mathbf{x}_j, \mathbf{x}_k)$ is the second-order interaction between $\mathbf{x}_j$ and $\mathbf{x}_k$, and so on. From Equation \ref{decomposition}, it is clear that the highest order of interaction is the maximum depth of a tree, $D$. The minimum sum of observation weights in a leaf determines the variance of $\hat{\beta}$'s, estimated weights of leaves in a tree. If the minimum sum is large, the growth of a tree will be conservative, which means fewer leaves will be grown and thus a smaller variance of $\hat{\beta}$'s. 

Finally, we discuss how Bayesian optimization \citep{snoek2012practical} is applied for tuning $\gamma$, $\lambda$, and $\alpha$ in Equation \ref{norm012}. We choose Bayesian optimization because brute-force search such as grid search or random search is time-consuming in a scenario where there are three hyperparameters. Here we treat hyperparameter optimization as a black-box process. We define a configuration of tuning hyperparameters, $\mathbf{\theta} = (\gamma, \lambda, \alpha)$, for the function we want to minimize,
\begin{equation} \label{cv_test_error}
    \text{CVTE}(\gamma, \lambda, \alpha | \mathbf{y}, \mathbf{X}) = \frac{1}{n} \sum_{i=1}^{n} (y_i - f^B(\mathbf{x}_i))^2,
\end{equation}
where $\text{CVTE}(\cdot)$ is a $K$-fold cross-validation error score, $\mathbf{x}_i$ is a row vector containing $p$ features of an observation, and $f^B(\cdot)$ is the fitted boosted tree model with $B$ individual trees. An early stopping criteria states that if adding new trees does not decrease cross-validation error within five trees, the boosting iteration should be stopped before the maximum number of trees is reached. The value of $B$ is equal to the number of trees in the sequence, where the combination of trees provides the lowest cross-validation error score. This error score is the output value from $\text{CVTE}(\cdot)$ in Equation \ref{cv_test_error}. The support sets of $\gamma$, $\lambda$, and $\alpha$ are located within a tuning region, $[0,20]$. The optimal combination is used in the final model for comparison. Details are provided in Section \ref{accu_consis}.

\subsection{Knockoff Variables in Boosted Tree Models} \label{subsec_knockoff}

\cite{barber2015controlling} proposed the knockoff filter, a new variable selection method that controls the FDR. The knockoff filter generates knockoff variables that mimic the correlation structure of original variables but are not associated with the response. These knockoff variables are used as controls for the original variables, so only the original variables that are highly associated with the response are selected. The knockoff filter has been shown to provide accurate FDR control, which cannot be realistically achieved using permutation methods. \cite{barber2015controlling}'s filter works under the following conditions, where for a fixed design, $\mathbf{X}_{n \times p}$, $n>p$, and $\mathbf{y}$ follows a linear Gaussian model. \cite{candes2018panning} introduced model-X knockoffs, which extended the model assumption to high-dimensional and covariate selection from random known distributions. Following the definition of model-X knockoffs in \cite{candes2018panning}, we restate the definition for boosted tree models here. 

\begin{definition}{} \label{knockoff}
For a row vector of $p$ random variables $\mathbf{x}_{1 \times p}= \left ( x_1, x_2, ..., x_p \right )$, where each $x_j$ is a random variable that represents a feature, the corresponding model-X knockoff variables $\mathbf{z}_{1 \times p}= \left ( z_1, z_2, ..., z_p \right )$ are constructed such that:
    \begin{enumerate}
        \item for a combined random vector $(\mathbf{x},\mathbf{z})_{1 \times 2p}$, if its $j$-th random variable is switched with its $(j+p)$-th random variable ($j=1,...,p$) (i.e, an original variable is switched with its knockoff counterpart), the distribution of the new random vector is invariant to $(\mathbf{x},\mathbf{z})_{1 \times 2p}$; and
        \item $\mathbf{z}_{1 \times p} \independent \mathbf{y} | \mathbf{x}_{1 \times p}$, where $\mathbf{y}$ is the response.
    \end{enumerate}
\end{definition}

Based on the definition, to achieve high-dimensional FDR-controlled variable selection, qualified knockoff variables should satisfy two properties: (1) distribution equality with original variables and (2) independence of response. Two sampling methods, namely exact constructions and approximate construction, have been introduced to generate knockoffs \citep{candes2018panning}. For exact construction, a new knockoff variable $z_j$ is sampled from the conditional distribution $L(x_j|x_{-j},z_{1:j-1})$, where $j=1,...,p$, and $x_{-j}=(x_{1:(j-1)},x_{(j+1):p})$. Approximate construction focuses on whether $(\mathbf{x},\mathbf{z})_{1 \times 2p}$ retains its first two moments of a distribution after swapping. Given this summary of the two methods, it is apparent that the approximate construction method requires less complex computations than the exact construction method.

In this article, we compare three algorithms for knockoff generation. The first algorithm, \textit{Approximate Construction} (AC), is available in the \textit{knockoff} R package \citep{candes2018panning}, where the covariance matrix of original variables, cov$(\bf{X}_{n \times p})$, is estimated directly. The estimated covariance matrix is shrunk to the identity matrix if it is not positive definite \citep{opgen2007accurate}. In addition, we propose two other algorithms: one that is similar to the AC algorithm described above and one without Gaussian assumption. The second algorithm (i.e., the one that is similar to the AC algorithm) is a Sparse Constructions (SC) algorithm. Instead of estimating the covariance matrix directly, we conduct sparse estimation of the covariance matrix \citep{bien2011sparse}, which generates a sparse matrix with a simpler structure. The rest of the algorithm is the same as the AC algorithm. The third algorithm (i.e., the one without Gaussian assumption) is a \textit{Principal Component Constructions} (PCC) algorithm, which is inspired by Algorithm A.1 in \cite{shen2019false}. In accordance with Definition \ref{knockoff}, we describe our proposed PCC algorithm in Algorithm \ref{pcc} below. Unlike the AC and SC algorithms, the PCC algorithm does not require data with a Gaussian assumption. 

\begin{algorithm}{\textit{Principal Component Construction} (PCC)} \label{pcc}
    
    For each column vector of original variable $\mathbf{x}_{j}$, where $j=1,...,p$,
    
    \begin{enumerate}
        \item Conduct principal component analysis on matrix $(\mathbf{X}_{-j},\mathbf{Z}_{1:j-1})_{n \times (p + j - 2)}$, where $\mathbf{X}_{-j}$ is matrix $\mathbf{X}$ without the $j$-th column and $\mathbf{Z}_{1:j-1}$ is the first $(j-1)$ columns in matrix $\mathbf{Z}$. When $j=1$, $\mathbf{Z}_{1:j-1}$ is empty.
        \item Denote $K$ as the number of principal components chosen for a regression model, $K = 1, ..., n-1$. For a fixed $K$, fit $\mathbf{x}_{j}$ on $K$ PCs. There is a tradeoff in that the larger the $K$, the more akin the knockoff will be to the original variables. This results in a smaller type 1 error but weaker power of the test.
        \item Compute a residual vector $\mathbf{\varepsilon}_{n}=(\mathbf{x}_{j}-\hat{\mathbf{x}}_{j})$.
        \item Permute $\mathbf{\varepsilon}_{n}$ randomly. Denote the permuted vector as $\mathbf{\varepsilon}^{*}_{n}$.
        \item Set $\mathbf{z}_{j}=\hat{\mathbf{x}}_{j}+\mathbf{\varepsilon}^{*}_{n}$, and combine it with the current knockoff matrix $\mathbf{Z}_{1:j-1}$. 
    \end{enumerate}
\end{algorithm}

This algorithm was designed in accordance with our definition of knockoff. Using linear regression models, the empirical conditional distribution of $\mathbf{x}_{j}$, $L(\mathbf{x}_{j} | \mathbf{X}_{-j},\mathbf{Z}_{1:j-1})$ can be estimated. Using permutation, we eliminate the Gaussian assumption. Meanwhile, the generated $\mathbf{z}_{j}$ is independent of the response $\mathbf{y}$, since $\mathbf{y}$ is ignored in our algorithm. Note that this is a sequential algorithm, so computation could be expensive. In \cite{shen2019false}, the residuals are assumed to be approximately independently and identically distributed. We keep this assumption for our algorithm.

To evaluate knockoffs generated by these methods, we propose the \textit{Mean Absolute Angle of Columns} (MAAC) metric for checking vector independence, and use the \textit{Kernel Maximum Mean Discrepancy} (KMMD) metric to test distribution similarity \citep{gretton2007kernel}.

\begin{definition}{\textit{Mean Absolute Angle of Columns} (MAAC)} \label{maac}

For two column vectors with the same length, $\mathbf{x}$ and $\mathbf{z}$, we define
\begin{equation}
    \text{MAAC}(\mathbf{x}, \mathbf{z}) = \text{arccos} \left| \frac{\mathbf{x}^\top\mathbf{z}}{||\mathbf{x}|| \cdot ||\mathbf{z}||} \right|.
\end{equation}

For two matrices with the same dimensions, $\mathbf{X}$ and $\mathbf{Z}$, with $p$ columns, we define
\begin{equation}
    \text{MAAC}(\mathbf{X}, \mathbf{Z}) = \frac{1}{p} \sum_{j=1}^{p} \text{arccos} \left| \frac{\mathbf{x}_{j}^\top\mathbf{z}_{j}}{||\mathbf{x}_{j}|| \cdot ||\mathbf{z}_{j}||} \right|.
\end{equation}
\end{definition}

MAAC indicates the correlation between corresponding columns in two matrices. For knockoff variable selection, we know that the weaker the correlation, the more powerful the test. The KMMD metric is used to perform a non-parametric distribution test. The null hypothesis test, $H_{0}$, is that the row vectors $\mathbf{x}_{i}$ in $\mathbf{X}$ and $\mathbf{z}_{i}$ in $\mathbf{Z}$ come from the same distribution. In summary, MAAC is for type 2 error control, and KMMD is for type 1 error control. In Section \ref{knockoff_property}, simulation tests are described, and permutation tests are used for comparison.

\subsection{Variable Importance Test Statistics in Tree Models} \label{sec:var_imp}

For each original and knockoff variable pair, \cite{barber2015controlling} suggested the \textit{Lasso signed max} (LSM) statistic as the test statistic for variable selection. \cite{candes2018panning} later proposed the \textit{Lasso coefficient difference} (LCD) statistic for the same purpose. While both of these test statistics were designed specifically for Lasso, there are also test statistics designed for tree models. Table \ref{var_imp} provides a list of common variable importance test statistics for boosted tree models. Given that different test statistics will provide different results for importance ranking, it is difficult to determine which test statistic is the best choice in a specific scenario.

\begin{table}
    \caption[Available variable importance test statistics in tree models]{Available variable importance test statistics in boosted tree models.}
    \centering
    \begin{tabular}{c|l}
    \midrule\midrule
    Test Statistic & Definition \\
    \midrule\midrule
    \multirow{2}{*}{Cover} & The number of times a variable is used, weighted by the \\
    & amount of training data in the corresponding nodes \\
    \hline
    Gain \citep{breiman1984classification} & The total loss reduction gained when using a variable \\
    \hline
    Saabas \citep{saabas2014interpreting} & The change in the model's expected outputs \\
    \hline
    \multirow{2}{*}{Tree SHAP \citep{lundberg2018consistent}} & The sum of weighted difference of conditional expectations \\
    & with and without a variable \\
    \hline
    Weight & The number of times a variable is used \\
    \midrule\midrule
    \end{tabular}
    \label{var_imp}
\end{table}

Among these statistics, SHAP, which is based on game theory and local explanations, has the demonstrated advantage of retaining both consistency and local accuracy \citep{lundberg2017unified}. \cite{lundberg2017consistent} introduced feature attribution for tree ensembles using the \textit{additive feature attribution methods} defined by \cite{lundberg2017unified}. They also proposed the Tree SHAP algorithm, which reduces the complexity of computing exact SHAP values from $\mathcal{O}(BL2^p)$ to $\mathcal{O}(BLD^2)$, where $B$ is the number of trees, $L=\max(|T_1|,...,|T_B|)$ is the maximum number of leaves in any tree, $p$ is the number of variables, and $D$ is the maximum depth of any tree. Further, \cite{lundberg2018consistent} extended SHAP values to SHAP interaction values. In \textit{additive feature attribution methods}, $f(\cdot)$ in Equation \ref{sum_functions} is the original model, and an explanation model, $h(\cdot)$, is defined for $f(\cdot)$ such that
\begin{equation}
    h(\mathbf{u}) = \phi_0 + \mathbf{\phi}^\top \mathbf{u},
\end{equation}
where $\mathbf{u} \in \{0, 1\}^p$ is a binary $p$ dimensional column vector, $u_j=1$ if the $j$-th variable is observed, $u_j=0$ if it is missing, and $\mathbf{\phi}$ stands for feature attribution values. The exact value of each $\phi_j$ can be calculated as
\begin{equation}
    \phi_j = \sum_{S \subseteq P \setminus \{j\}} \frac{|S|!(p-|S|-1)!}{p!}[f_x(S\cup \{j\})-f_x(S)],
\end{equation}
where $P$ is the set of all variables, $S$ is a subset of $P$ with non-zero indices in $\mathbf{u}$, and $f_x(S)=E[f(X)|X_S]$ is defined as a conditional expectation. This $\phi_j$ is the SHAP value for the $j$-th variable.

Following the concept outlined by \cite{barber2015controlling} and \cite{candes2018panning}, we move on to define a test statistic for knockoff variables in tree models that can control the FDR. For simplicity, we use the SHAP value as an example, but other test statistics can be applied to the knockoff variable using a similar procedure. We denote a pair of original and knockoff variables as $\mathbf{x}_j$ and $\mathbf{z}_j$, which is the same notation used in Section \ref{subsec_knockoff}. Accordingly, their respective SHAP values are $\phi_j^x$ and $\phi_j^z$. The hypothesis test in which we are interested is:
\begin{equation*}
    \text{$H_0$: The original variable, $\mathbf{x}_j$, and the knockoff variable, $\mathbf{z}_j$, are equally associated with response;}
\end{equation*}
\begin{equation*}
    \text{$H_a$: The original variable, $\mathbf{x}_j$, is more closely associated with response than the knockoff variable, $\mathbf{z}_j$.}
\end{equation*}
Here, the \textit{Shapley explanation absolute difference} (SEAD) statistic is defined as
\begin{equation}
    T_j = |\phi_j^x(\gamma, \lambda, \alpha)| - |\phi_j^z(\gamma, \lambda, \alpha)|,
\end{equation}
where $\gamma$, $\lambda$, and $\alpha$ are hyperparameters used in Equation \ref{cv_test_error}. A larger positive $T_j$ indicates that the response is more likely to be associated with the original variable than its knockoff. Under the null hypothesis, $T_j$ is equally likely to be positive or negative. Given that null $T_j$s are symmetric, for any fixed $t>0$, the false discovery proportion is
\begin{equation}
    \text{FDP}(t) = \frac{\#\{\text{null } j: T_j \geq t\}}{\#\{j: T_j \geq t\}}.
\end{equation}
This can be estimated by
\begin{equation}
    \hat{\text{FDP}}(t) = \min \left\{ \frac{\#\{ j: T_j \leq -t\}}{\#\{j: T_j \geq t\}}, 1 \right\},
\end{equation}
where $\hat{\text{FDP}}(t) = 1$ when there are more $T_j$s below $-t$ than above $t$. \cite{candes2018panning} provided equations for both the modified FDR and the typical FDR. Here, we accept the latter as it is more conservative. A threshold $\tau$ is defined such that
\begin{equation} \label{fdr}
    \tau(t) = \min \left\{ \frac{\#\{ j: T_j \leq -t\}+1}{\#\{j: T_j \geq t\}} \leq \delta \right\},
\end{equation}
where $\delta$ is the level under which we would like to keep the FDR. A set of selected variables is determined as
\begin{equation}
    \hat{S}(\tau) = \{j: T_j \geq \tau \}.
\end{equation}
Then, the FDR is controlled as
\begin{equation}
    E\left( \frac{| \hat{S} \cap S_0 |}{|\hat{S}| \vee 1} \right) \leq \delta,
\end{equation}
where $S_0 \subseteq \{1,2,...,p\}$ is a set of noise variables. The performance of the SEAD statistic in various models is evaluated in the following sections, and test statistics constructed using other methods are tested for comparison.

\subsection{Knockoff Boosted Tree Algorithm}

\begin{figure}[ht]
  \centering
  \includegraphics[scale=0.55]{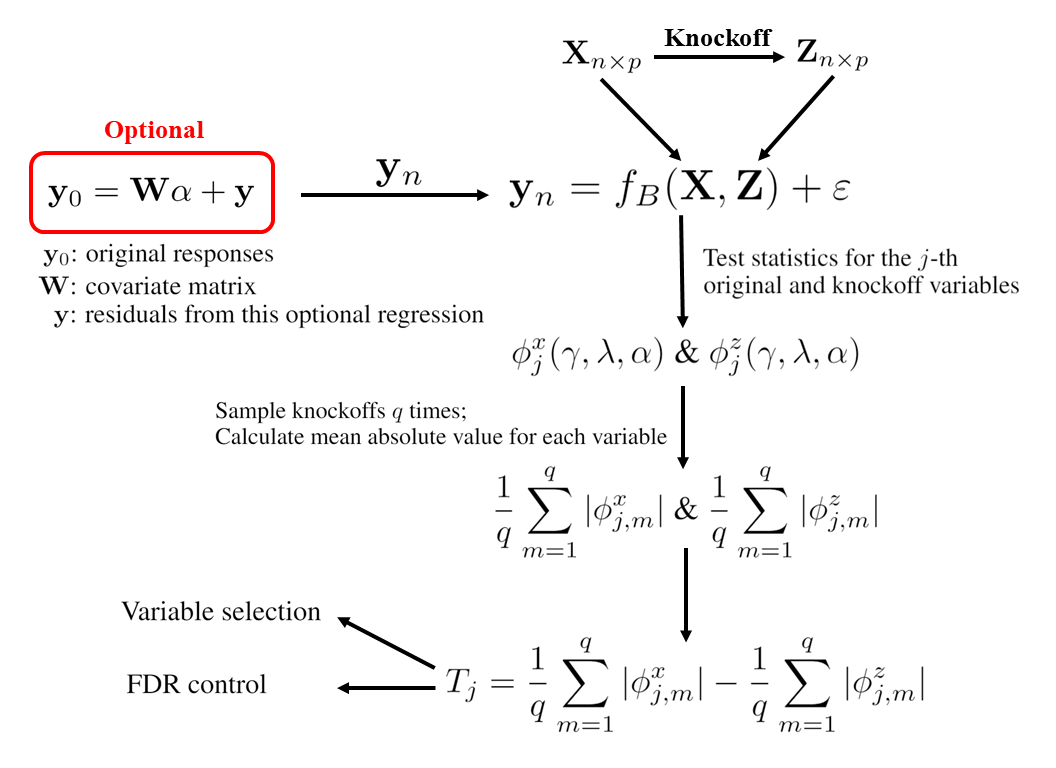}
  \caption[Knockoff boosted tree flowchart]{Knockoff boosted tree flowchart. An optional step is circled in red.}
  \label{fig:kobt_flow}
\end{figure}

To perform model-free variable selection with FDR control, we propose a robust multiple-stage KnockOff Boosted Tree (KOBT) algorithm. Figure \ref{fig:kobt_flow} is a flowchart that outlines details of the procedure. The algorithm contains four main steps and one optional step (circled in red): (1) sample knockoff matrix $\mathbf{Z}$ according to original matrix $\mathbf{X}$; (2) grow boosted trees using combined predictors $(\mathbf{X}, \mathbf{Z})$; (3) calculate the test statistic, $\phi$, for each variable from the fitted boosted tree model,  $f_B(\mathbf{X},\mathbf{Z})$; and (4) conduct steps (1) to (3) $q$ times and get $T_j = \frac{1}{q}\sum_{m=1}^{q} |\phi_{j,m}^x| - \frac{1}{q}\sum_{m=1}^{q} |\phi_{j,m}^z|$ for the $j$-th original variable. A larger positive $T_j$ indicates a variable is more closely associated with response. The optional step is taken in the scenario in which some variables (shown as $\mathbf{W}$ in Figure \ref{fig:kobt_flow}) need to be retained in the final model. 

The first part of the algorithm is optional, and a regression model is added before the boosted tree model. Given that we want to retain some covariate variables in our final model, we conduct a regression, 
\begin{equation}
    \mathbf{y}_0 = \mathbf{W}\alpha + \mathbf{y},
\end{equation}
where $\mathbf{y}_0$ is the original responses, $\mathbf{W}$ is the covariate matrix, and $\mathbf{y}$ stands for residuals from this optional regression. This $\mathbf{y}$ is treated as new responses for the subsequent boosted tree fitting.

For the first required step of KOBT, we generate knockoff variables $\mathbf{Z}$ conditional on original variables $\mathbf{X}$. With the assumption that the data in $\mathbf{X}$ follow a Gaussian distribution, the column mean and covariance matrix of $\mathbf{X}$ are estimated. As discussed in Section \ref{subsec_knockoff}, there are two choices for covariance matrix estimation: shrinking the estimation to the identity matrix \citep{opgen2007accurate} and sparse estimation \citep{bien2011sparse}. Without a Gaussian assumption, we propose a strategy based on principal components and permutation (see Section \ref{subsec_knockoff} for details). The properties of knockoff variables are compared in Section \ref{knockoff_property}.

During the process of model fitting, hyperparameters $\gamma$, $\lambda$, and $\alpha$ are tuned through Bayesian optimization. Once a boosted model is fitted, test statistics showing importance are calculated. Multiple statistics are considered, such as gain, cover, weight, SHAP value, and Saabas value. Their values are discussed in Section \ref{imp_rank}. The above steps, from generating knockoff variables to calculating test statistics, are repeated $q$ times to achieve the mean absolute value for each variable, where $q$ should be sufficiently large. According to the strategy of applying knockoff variables \citep{barber2015controlling, candes2018panning}, for each pair of original and knockoff variables, a test statistic is calculated,
\begin{equation} \label{sead}
    T_j = \frac{1}{q}\sum_{m=1}^{q} |\phi_{j,m}^x| - \frac{1}{q}\sum_{m=1}^{q} |\phi_{j,m}^z|,
\end{equation}
which is the difference between the mean absolute test statistic values of each variable pair. It is straightforward that a larger test statistic proves that the original variable is more important than its corresponding knockoff. Finally, the FDR of variable selection is controlled through values of all generated $T$s, as shown in Equation \ref{fdr}.

We use simulation tests to address the following topics: (a) control of type I and type II errors for knockoffs generated by different methods; (b) the properties of different variable importance statistics in boosted tree models; and (c) power and false discovery control using various combinations of knockoffs and statistics. Further, the performance of boosted tree fitting for different model structures is compared.

\section{Simulation Studies}
\label{sec:simulation}

\subsection{Incorrect but Related Variables in Boosted Tree Models} \label{accu_consis}

First, we describe the models and the boosted tree framework used in the following simulation tests. For the design matrix $\mathbf{X}_{n \times p}^0$, we simulate $n=200$ samples and $p=1000$ variables. Each row vector in design matrix $\mathbf{X}_{n \times p}^0$ is generated as $\mathbf{x}_i \sim N_p (\mathbf{0}, \mathbf{\Sigma})$ with block dependence structure matrix $\mathbf{\Sigma}=\text{diag}(\Sigma_{1}, \Sigma_{2}, \Sigma_{3}, ...)$, where each $\Sigma_{i}$ is a $\pi p \times \pi p$ matrix with matrix element $\sigma_{j,k} =(\rho^{|j-k|})$. We set $\pi = 0.01$ and $\rho = 0.1$. Because we want to test model-free variable selection, we generate the response as
\begin{equation}
    \mathbf{y}=\mathbf{X}\mathbf{\beta} + \mathbf{\varepsilon}, 
\end{equation}
where $\mathbf{\beta}=(\mathbf{\beta}_{\pi p}^\top, \mathbf{0}_{p-\pi p}^\top)^\top$, $\beta_{\pi p}$ is a vector with equal elements, $\mathbf{\varepsilon} \sim N_p(\mathbf{0}, \mathbf{I}_p) $, and $\mathbf{X}$ is transformed from initial $\mathbf{X}^0$ with four different structures, namely:
\begin{enumerate}
    \item Main-effect model: $\mathbf{X}=\mathbf{X}^0$.
    \item Interaction effect model: each $x_{i,j} = x_{2i-1,j}^0 x_{2i,j}^0$, for $i=1,...,\pi p/2$. This means we keep $\pi p/2$ signal variables the same as the main-effect models.
    \item Exponential effect model: each $x_{i,j} = \exp{(x_{i,j}^0)}$, for $i=1,...,\pi p$.
    \item Quadratic effect model: each $x_{i,j} = (x_{i,j}^0)^2$, for $i=1,...,\pi p$.
\end{enumerate}
While we generate the response $\mathbf{y}$ using the transformed $\mathbf{X}$, we work under the notion of having no information about the transformed $\mathbf{X}$ and fit the models with the initial design matrix $\mathbf{X}^0$. Thus, we use incorrect but related variables for fitting boosted tree models. While this is challenging, there is no guarantee that the correct variables will be used for real data modeling, so it is possible to see boosted tree modeling performance with incorrect but related variables.

After defining the simulated data values, we move on to our boosted tree models. The booster used for growing our tree are the gradient boosted tree (GBRT in Table \ref{test_error}) and dropouts meet multiple additive regression tree (DART in Table \ref{test_error}) \citep{rashmi2015dart}. As suggested in \cite{friedman2000additive} and \cite{zhang2005boosting}, a small learning rate is preferred in boosted model fitting. For our simulation, we fix the learning rate $\gamma = 0.01$. In this scenario, the trees previously added to the boosted model are more important. This overfits the model with trees added earlier. In iterations of DART model training, some trees are dropped randomly to avoid overfitting while in traditional GBRT, all trees are kept once they join a model. We compare these two algorithms in the following simulations.

The tuning of hyperparameters is tricky, especially for a complex system such as a boosted tree with many parameters. The purpose of our simulation is to compare the performance of boosted trees under a series of conditions and thus we choose $10$ initial hyperparameter values of $(\gamma, \lambda, \alpha)$ and conduct Bayesian optimization iterations $20$ times for each fitting. We use the average $10$-fold cross-validation error score for model evaluation. An early stopping criteria is set: if the average $10$-fold cross-validation error score has not improved after five training iterations, no new trees are added. The maximum depth of each tree is fixed at $2$, $3$, $4$, $5$, and $6$. The minimum weight of each single leaf is $10$. Note that 
\begin{enumerate}
    \item We can increase the number of initial $(\gamma, \lambda, \alpha)$ hyperparameter values and Bayesian optimization iterations to find new $(\gamma, \lambda, \alpha)$ that can improve the performance of the final model. However, this is not within the scope of our simulation tests.
    \item For real data, we can choose a larger number for the maximum depth of a boosted tree model. Here, we use only $2$ to $6$ because this is a simple simulation test.
\end{enumerate}

\begin{figure}[!htbp]
  \centering
  \includegraphics[scale=0.54]{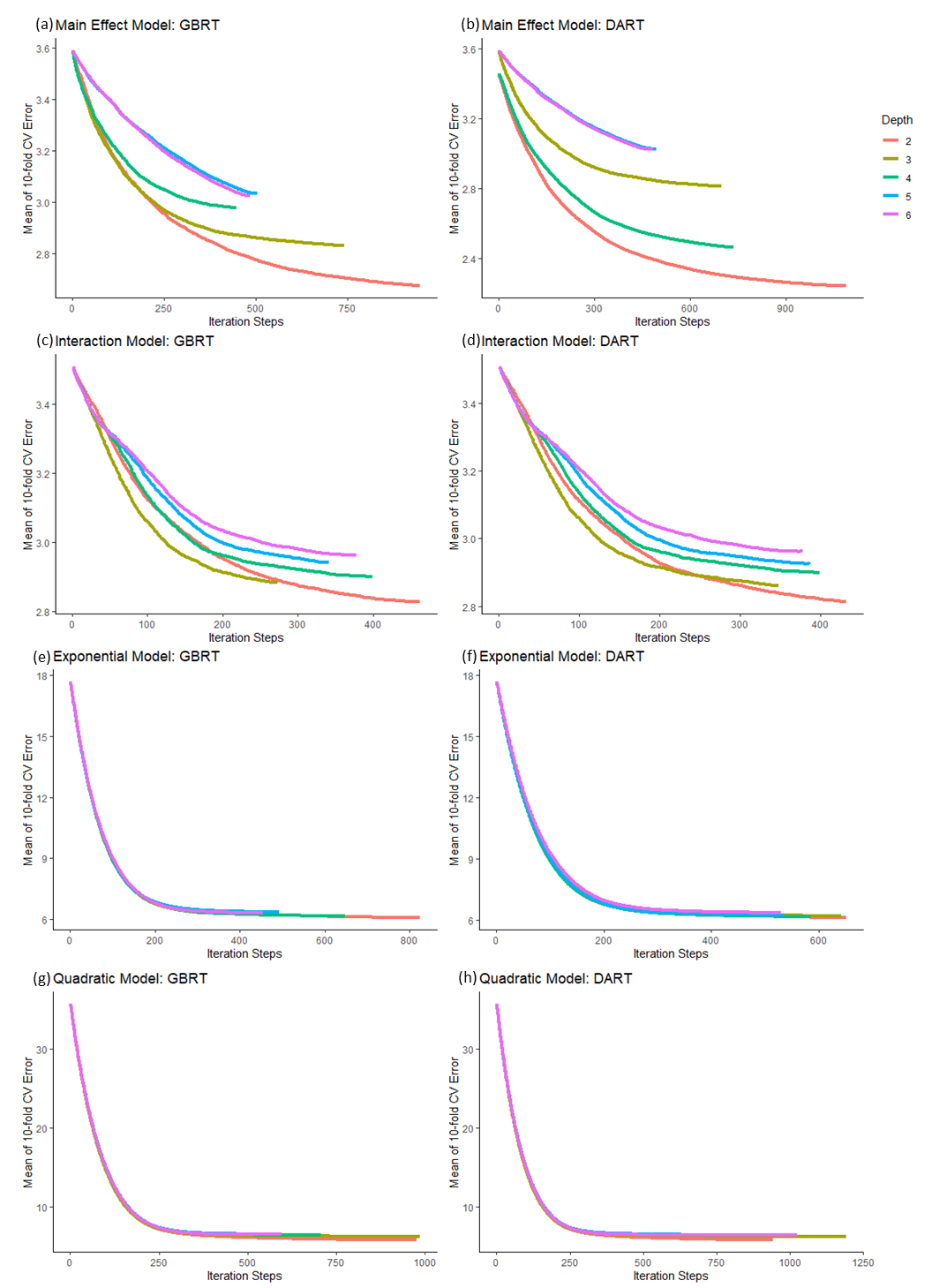}
  \caption[Examples of boosted tree iteration steps for all models and boosters.]{Examples of boosted tree iteration steps for all models and boosters. (a) Main-effect model with GBRT; (b) Main-effect model with DART; (c) Interaction model with GBRT; (d) Interaction model with DART; (e) Exponential model with GBRT; (f) Exponential model with DART; (g) Quadratic model with GBRT; and (h) Quadratic model with DART.}
  \label{fig:test_error_line_plot}
\end{figure}

For each combination of the above four models, two boosters, and five maximum tree depths, 100 boosted tree models are fitted. Table \ref{test_error} lists the means and standard errors of 100 10-fold cross-validation error scores. We use the same design matrices $\mathbf{X}_{1}^0, ..., \mathbf{X}_{100}^0$ for generating transformed matrices. For each model structure, $\mathbf{X}_{1}, ..., \mathbf{X}_{100}$ are created according to our definitions. Among all four model structures, the main-effect model, which is the correct variable model, has the lowest cross-validation prediction error. The incorrect but related models, ranked from the lowest to highest cross-validation prediction error, are interaction, exponential, and then quadratic. This is reasonable since the structure of tree models is formed naturally by interaction among variables. From the standard errors of means, we conclude that predictions for the main-effect and interaction models are more stable than predictions for the exponential and quadratic models. The results indicate that although prediction ability differs by the various true models, it is realistic to use incorrect but related variables in boosted tree modeling. As for maximum tree depth, since the correlation between two variables is chosen as $0.1^d$, where $d$ is the index difference of each pair of variables, the correlation decreases rapidly as we choose two variables that are farther away from each other. This is why, in Table \ref{test_error}, the cross-validation error score is the lowest when depth $=2$. This implies that prior knowledge of the strength of variable correlation is important, as we need to choose an appropriate value for tree depth. Finally, under the conditions of this simulation, there is no obvious difference between GBRT and DART boosters. Figure \ref{fig:test_error_line_plot} shows examples of boosted tree iteration steps for all models and boosters. 

\begin{table}
    \caption[Means and standard errors of 100 10-fold cross-validation error scores]{Means and standard errors of 100 10-fold cross-validation error scores, mean (SD), where sample size $n = 200$, feature size $p = 1000$, signal proportion $\pi = 0.01$, signal strength $\beta = 1$, learning rate $\eta = 0.01$, minimum child weight $w_{min} = 10$, and all samples and features are considered in each iteration step.}
    \centering
    \begin{tabular}{cc|ccccc}
    \midrule\midrule
    Models & Boosters & Max Depth $= 2$ & 3 & 4 & 5 & 6 \\
    \midrule\midrule
    \multirow{2}{*}{Main Effect} & GBRT & 2.394 (0.012) & 2.590 (0.012) & 2.709 (0.014) & 2.741 (0.014) & 2.743 (0.014) \\
    & DART & 2.389 (0.012) & 2.591 (0.013) & 2.710 (0.013) & 2.741 (0.014) & 2.743 (0.014) \\
    \hline
    \multirow{2}{*}{Interaction} & GBRT & 2.853 (0.018) & 2.862 (0.018) & 2.878 (0.018) & 2.883 (0.018) & 2.887 (0.018) \\
    & DART & 2.852 (0.018) & 2.861 (0.018) & 2.877 (0.018) & 2.882 (0.018) & 2.861 (0.023) \\
    \hline
    \multirow{2}{*}{Exponential} & GBRT & 5.346 (0.059) & 5.369 (0.055) & 5.446 (0.054) & 5.461 (0.058) & 5.551 (0.057) \\
    & DART & 5.344 (0.058) & 5.350 (0.052) & 5.461 (0.056) & 5.512 (0.056) & 5.548 (0.057) \\
    \hline
    \multirow{2}{*}{Quadratic} & GBRT & 5.704 (0.033) & 6.063 (0.032) & 6.269 (0.034) & 6.366 (0.034) & 6.424 (0.035) \\
    & DART & 5.713 (0.034) & 6.059 (0.032) & 6.275 (0.033) & 6.368 (0.034) & 6.401 (0.034) \\
    \midrule\midrule
    \end{tabular}
    \label{test_error}
\end{table}

\subsection{Power and Type I Error Control of Knockoff Variables} \label{knockoff_property}

As proposed in Section \ref{subsec_knockoff}, there are three strategies to generate knockoff variables: Gaussian assumption with shrunk covariance matrix, Gaussian assumption with sparse covariance matrix, and permuted residuals from principal component regression. We are interested in evaluating the power and type I error performance using different knockoff variables for testing. We apply the MAAC metric, which is defined in Section \ref{subsec_knockoff}, for power evaluation. A larger positive MAAC value shows that two tested matrices have less related column space, so the difference between a real signal and its knockoff is more significant. As for type I errors, we use KMMD to test if row vectors in the original design matrix $\mathbf{X}$ and knockoff matrix $\mathbf{Z}$ are drawn from the same distribution. A larger positive test statistic implies that it is more likely to reject the null hypothesis, which assumes that these two row vectors are from the same distribution. If knockoff variables are highly different from the original variables, this leads to large type I errors. In other words, we want knockoff variables to be drawn from the same distribution as the original variables, but with sufficiently different values. As always, there is a tradeoff between power and type I error control.

\begin{figure}[!htbp]
  \centering
  \includegraphics[scale=0.55]{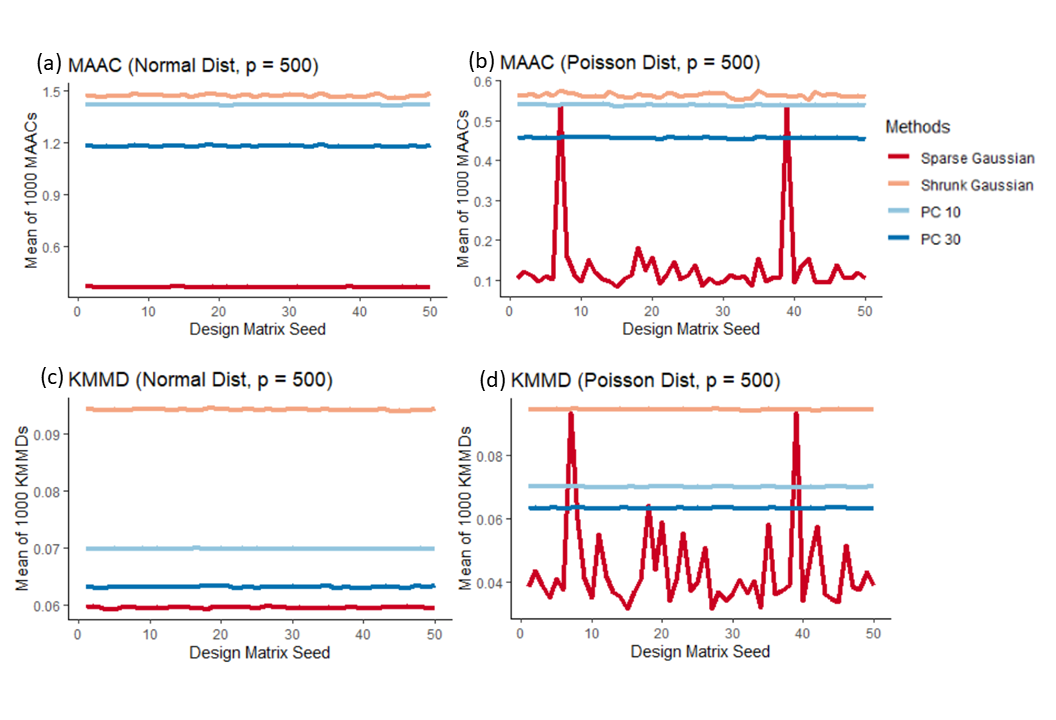}
  \caption[MAAC and KMMD of knockoffs with Normal and Poisson original variables.]{MAAC and KMMD of knockoffs with Normal and Poisson original variables. (a) MAAC with normally distributed $\mathbf{X}$; (b) MAAC with Poisson-distributed $\mathbf{X}$; (c) KMMD with normally distributed $\mathbf{X}$; and (d) KMMD with Poisson-distributed $\mathbf{X}$.}
  \label{fig:evaluation_knockoff}
\end{figure}

We first assume the row vector in design matrix $\mathbf{X}$ is from a normal distribution. We simulate $n=100$ samples and $p=500$ variables. Each row vector in design matrix $\mathbf{X}_{n \times p}$ is generated as $\mathbf{x}_i \sim N_p (\mathbf{0}, \mathbf{\Sigma})$ with block dependence structure matrix $\mathbf{\Sigma}=\text{diag}(\Sigma_{1}, \Sigma_{2}, \Sigma_{3}, ...)$, where each $\Sigma_{i}$ is a $\pi p \times \pi p$ matrix with matrix element $\sigma_{j,k} =(\rho^{|j-k|})$. We set $\pi = 0.01$ and $\rho = 0.1$. In Figure \ref{fig:evaluation_knockoff}, we simulate $50$ original design matrices and generate $1,000$ corresponding knockoff matrices for each original design matrix. Four kinds of knockoffs are created for comparison: shrunk Gaussian, sparse Gaussian, permuted principal component $10$, and permuted principal component $30$. The $y$-axis is calculated as the mean value of each $1,000$ knockoff samples from one original matrix. The $x$-axis is the seed used to sample the original matrix. Therefore, $y$ values of different methods at the same $x$ location are comparable. It is clear from Figure \ref{fig:evaluation_knockoff} (a) and (c) that all methods are consistent under normal assumption. The MAAC values in Figure \ref{fig:evaluation_knockoff} (a) show that the sparse method provides the most similar knockoffs, which also means they are the least powerful. On the other hand, the shrunk method has the most different knockoffs. Principal component methods provide knockoffs in between. This makes sense since the shrunk Gaussian method directly estimate the covariance matrix of $\mathbf{X}$ and shrinks it to the identity matrix if it is not definitely positive, while the sparse Gaussian method has zeros in the estimated matrix given its sparse assumption. In Figure \ref{fig:evaluation_knockoff} (c), the KMMD plot has the same order as MAAC. However, the sparse method has the lowest type I error while the shrunk method has the highest. 

Since both the shrunk Gaussian and sparse Gaussian methods assume $\mathbf{X}$ follows a normal distribution, it is interesting to see what would happen if a design matrix does not follow a normal distribution. Figure \ref{fig:evaluation_knockoff} (b) and (d) show that the sparse Gaussian method is not as consistent for a Poisson distribution as it is for normal distribution. The principal component method is consistent, however, because it does not depend on the normal assumption. Despite some exceptions, for most design matrices, the $y$-axis order of these four curves is the same as for normal distribution. In summary, the sparse Gaussian method is the most conservative, the shrunk Gaussian method is the most powerful, and the principal component method lies in between. When we increase the number of principal components, the performance of the generated knockoff is similar to the performance of the shrunk method.

\subsection{Comparison of Variable Importance Ranking Statistics} \label{imp_rank}

For each variable in a tree model, feature importance can be used to show its importance in the model. As discussed in Section \ref{sec:var_imp}, a few test statistics can represent this importance. In this section, we conduct simulation tests comparing \textit{gain}, \textit{cover}, \textit{frequency}, \textit{SHAP}, and \textit{Saabas}. Since we are only interested in the ranking ability of these test statistics for selected variables, not the performance of variable selection, which is conducted by tree growing itself, we define a score called the ranking ratio (RR) as
\begin{equation}
    \text{RR} = \frac{\#\text{ Noise variables ranked above the last signal in ranking}}{\#\text{ Noise variables}}.
\end{equation}
The RR contains two parts: the number of noise variables selected by the boosted tree and ranked above the lowest ranked signal among all selected signals, and the total number of mis-selected noise variables. This score, which ranges within $(0,1)$, indicates how many false positive decisions must be made, if, for a given set of variables chosen during tree growing, we want to include all signals for variable selection. The score is undefined if no signal or noise are selected in a tree model. It evaluates the ranking ability of each variable importance ranking statistic. A small RR score indicates better ranking ability of variable importance.

We keep the model described in Section \ref{knockoff_property} and simulate $n=100$ samples and $p=500$ variables. Each row vector in design matrix $\mathbf{X}_{n \times p}$ is generated as $\mathbf{x}_i \sim N_p (\mathbf{0}, \mathbf{\Sigma})$, with block dependence structure matrix $\mathbf{\Sigma}=\text{diag}(\Sigma_{1}, \Sigma_{2}, \Sigma_{3}, ...)$, where each $\Sigma_{i}$ is a $\pi p \times \pi p$ matrix with matrix element $\sigma_{j,k} =(\rho^{|j-k|})$. We set $\pi = 0.01$ and $\rho = 0.1$. Following tree model hyperparameters in Section \ref{accu_consis}, the maximum depth of each tree is fixed at $6$. We use two boosters, GBRT and DART, four model structures, and two signal strengths. In Table \ref{main_ranking}, the correct variables are used for modeling. Each mean and standard error are calculated from $1,000$ simulations. In each signal strength-booster combination, means for the test statistic are close, and the standard errors are almost the same. This indicates these variable importance statistics have similar consistency even under different signal strengths and boosters. There is no obvious difference between the two boosters. Among the five variable importance statistics, \textit{frequency} has the weakest ranking ability given its highest RR score, regardless of strength or booster. In Tables \ref{inter_ranking}, \ref{exponential_ranking}, and \ref{2ndorder_ranking}, incorrect but related variables are used for modeling. For these models, \textit{cover} has the best performance as it returns the lowest RR scores in most cases. Note that we are only comparing five variable importance statistics under each combination of model structure, strength, and booster (i.e., each row in Tables \ref{main_ranking} to \ref{2ndorder_ranking}). Different rows have different boosted tree models and are thus incomparable. 

\begin{table}
    \caption[Mean and standard error of signal percentage and ranking ratio for test statistics in main effect models]{Mean and standard error of signal percentage and ranking ratio (RR) from $1,000$ simulation tests for ranking test statistics in \textbf{main-effect} models, where $n=100$, $p=500$.}
    \centering
    \begin{tabular}{cc|ccccc}
    \midrule\midrule
    Strength & Booster & Gain & Cover & Frequency & SHAP & Saabas \\
    \midrule\midrule
    \multirow{2}{*}{0.5} & GBRT & $0.705$ $(0.008)$ & $0.685$ $(0.008)$ & $0.716$ $(0.008)$ & $0.699$ $(0.008)$ & $0.700$ $(0.008)$ \\
    & DART & $0.714$ $(0.009)$ & $0.700$ $(0.010)$ & $0.729$ $(0.009)$ & $0.708$ $(0.010)$ & $0.711$ $(0.010)$ \\
    \hline
    \multirow{2}{*}{1.5} & GBRT & $0.699$ $(0.008)$ & $0.697$ $(0.008)$ & $0.723$ $(0.008)$ & $0.700$ $(0.008)$ & $0.700$ $(0.008)$ \\
    & DART & $0.694$ $(0.010)$ & $0.693$ $(0.010)$ & $0.713$ $(0.010)$ & $0.700$ $(0.010)$ & $0.703$ $(0.010)$ \\
    \midrule\midrule
    \end{tabular}
    \label{main_ranking}
\end{table}

\begin{table}
    \caption[Mean and standard error of signal percentage and ranking ratio for test statistics in interaction models]{Mean and standard error of signal percentage and ranking ratio (RR) from $1,000$ simulation tests for ranking test statistics in \textbf{interaction} models, where $n=100$, $p=500$.}
    \centering
    \begin{tabular}{cc|ccccc}
    \midrule\midrule
    Strength & Booster & Gain & Cover & Frequency & SHAP & Saabas \\
    \midrule\midrule
    \multirow{2}{*}{0.5} & GBRT & $0.564$ $(0.010)$ & $0.561$ $(0.010)$ & $0.586$ $(0.010)$ & $0.566$ $(0.010)$ & $0.565$ $(0.010)$ \\
    & DART & $0.562$ $(0.009)$ & $0.550$ $(0.010)$ & $0.581$ $(0.009)$ & $0.561$ $(0.010)$ & $0.562$ $(0.010)$ \\
    \hline
    \multirow{2}{*}{1.5} & GBRT & $0.626$ $(0.009)$ & $0.634$ $(0.009)$ & $0.654$ $(0.009)$ & $0.633$ $(0.009)$ & $0.635$ $(0.009)$ \\
    & DART & $0.625$ $(0.010)$ & $0.627$ $(0.010)$ & $0.649$ $(0.010)$ & $0.630$ $(0.010)$ & $0.632$ $(0.010)$ \\
    \midrule\midrule
    \end{tabular}
    \label{inter_ranking}
\end{table}

\begin{table}
    \caption[Mean and standard error of signal percentage and ranking ratio for test statistics in exponential models]{Mean and standard error of signal percentage and ranking ratio (RR) from $1,000$ simulation tests for ranking test statistics in \textbf{exponential} models, where $n=100$, $p=500$.}
    \centering
    \begin{tabular}{cc|ccccc}
    \midrule\midrule
    Strength & Booster & Gain & Cover & Frequency & SHAP & Saabas \\
    \midrule\midrule
    \multirow{2}{*}{0.5} & GBRT & $0.802$ $(0.006)$ & $0.783$ $(0.006)$ & $0.800$ $(0.006)$ & $0.799$ $(0.006)$ & $0.799$ $(0.006)$ \\
    & DART & $0.810$ $(0.006)$ & $0.784$ $(0.006)$ & $0.810$ $(0.006)$ & $0.808$ $(0.006)$ & $0.808$ $(0.006)$ \\
    \hline
    \multirow{2}{*}{1.5} & GBRT & $0.817$ $(0.005)$ & $0.809$ $(0.005)$ & $0.818$ $(0.005)$ & $0.821$ $(0.005)$ & $0.821$ $(0.005)$ \\
    & DART & $0.817$ $(0.005)$ & $0.808$ $(0.005)$ & $0.818$ $(0.005)$ & $0.820$ $(0.005)$ & $0.821$ $(0.005)$ \\
    \midrule\midrule
    \end{tabular}
    \label{exponential_ranking}
\end{table}

\begin{table}
    \caption[Mean and standard error of signal percentage and ranking ratio for test statistics in second order models]{Mean and standard error of signal percentage and ranking ratio (RR) from $1,000$ simulation tests for ranking test statistics in \textbf{second-order} models, where $n=100$, $p=500$.}
    \centering
    \begin{tabular}{cc|ccccc}
    \midrule\midrule
    Strength & Booster & Gain & Cover & Frequency & SHAP & Saabas \\
    \midrule\midrule
    \multirow{2}{*}{0.5} & GBRT & $0.784$ $(0.007)$ & $0.759$ $(0.007)$ & $0.762$ $(0.007)$ & $0.786$ $(0.007)$ & $0.786$ $(0.007)$ \\
    & DART & $0.772$ $(0.009)$ & $0.747$ $(0.010)$ & $0.749$ $(0.010)$ & $0.768$ $(0.009)$ & $0.768$ $(0.009)$ \\
    \hline
    \multirow{2}{*}{1.5} & GBRT & $0.832$ $(0.005)$ & $0.834$ $(0.005)$ & $0.823$ $(0.005)$ & $0.846$ $(0.005)$ & $0.846$ $(0.005)$ \\
    & DART & $0.834$ $(0.005)$ & $0.828$ $(0.005)$ & $0.821$ $(0.006)$ & $0.839$ $(0.005)$ & $0.840$ $(0.005)$ \\
    \midrule\midrule
    \end{tabular}
    \label{2ndorder_ranking}
\end{table}

\subsection{SEAD, Other Knockoff Test Statistics, and Their False Discovery Control}

In this final subsection describing the simulation, we combine all of the previous steps and perform the whole framework of the knockoff boosted tree (KOBT) algorithm. We are interested in the power and FDR of different combinations of ranking statistics and knockoff types. We simulate $n=100$ samples and $p=500$ variables. Each row vector in design matrix $\mathbf{X}_{n \times p}$ is generated as $\mathbf{x}_i \sim N_p (\mathbf{0}, \mathbf{\Sigma})$ or $\mathbf{x}_i \sim \text{Poisson}_p (\mathbf{5}, \mathbf{\Sigma})$ with block dependence structure matrix $\mathbf{\Sigma}=\text{diag}(\Sigma_{1}, \Sigma_{2}, \Sigma_{3}, ...)$, where each $\Sigma_{i}$ is a $\pi p \times \pi p$ matrix with matrix element $\sigma_{j,k} =(\rho^{|j-k|})$. We set $\pi = 0.01$ and $\rho = 0.1$. The maximum depth of each tree is fixed at $6$. For simplicity, we choose GBRT as the booster and the main-effect model as the true model. In Tables \ref{kobt_norm_power} to \ref{kobt_pois_fdr}, power and FDR for different combinations of knockoff types and ranking statistics are listed for comparison. We choose a condition where signals are sparse and weak so that the performance of each method is considerably distinct.

In Table \ref{kobt_norm_power}, $50$ normally distributed design matrices are simulated, each with $1,000$ shrunk Gaussian knockoffs, $1,000$ sparse Gaussian knockoffs, $1,000$ $10$-principal component knockoffs, and $1,000$ $30$-principal component knockoffs. We set the targeted FDR as $0.1$. The means and corresponding standard errors of power for each combination are listed. Among these four types of knockoffs, the shrunk Gaussian knockoff has the highest power, which is the same conclusion reached in Figure \ref{fig:evaluation_knockoff} (a). This is followed by $10$- and $30$-principal component knockoffs. The sparse Gaussian knockoff has the lowest power. Among the five importance statistics we test, \textit{frequency} is always the most powerful statistic for all knockoffs, while \textit{gain} is the least powerful statistic. The other three are moderately powerful and fall between \textit{frequency} and \textit{gain}. Table \ref{kobt_norm_fdr} shows the means and corresponding standard errors of the FDR for each combination. Since the targeted FDR is $0.1$, it is clear that the sparse Gaussian knockoff is the most conservative method and is the only one that can ensure the FDR stays under the targeted level. At the same order of power, the shrunk Gaussian knockoff has the highest FDR, followed by the two principal component knockoffs.

Table \ref{kobt_pois_power} shows the power of the statistics when the normality assumption is invalid. Fifty Poisson-distributed design matrices are simulated, each with $1,000$ shrunk Gaussian knockoffs, $1,000$ sparse Gaussian knockoffs, $1,000$ $10$-principal component knockoffs, and $1,000$ $30$-principal component knockoffs. Again, the targeted FDR is set as $0.1$. The power of the knockoffs is the same as for normal distribution. For the importance statistics, the order for Poisson-distributed matrices is different from that for normally distributed matrices, where \textit{cover} has the highest power, followed by \textit{frequency} and then the other three. Both the sparse Gaussian and $30$-principal component knockoffs can control the FDR to stay close to the targeted level. The shrunk Gaussian and $10$-principal component knockoffs have higher FDRs. The order of importance statistics for FDR is the same as their power ranking. In summary, the ranges in tables below show the trends of knockoffs and importance statistics.
\begin{table}[h!]
    \centering
    \begin{tabular}{|c|c|c|c|}
    \hline
    \textbf{More Conservative} & & & \textbf{More Powerful} \\
    \hline
    Sparse Gaussian & 30-Principal Component & 10-Principal Component & Shrunk Gaussian \\
    \hline
    \end{tabular}
\end{table}
\begin{table}[h!]
    \centering
    \begin{tabular}{|c||c|c|c|}
    \hline
    & \textbf{More Conservative} & & \textbf{More Powerful} \\
    \hline
    Normal & Gain & Cover, Saabas, and SHAP & Frequency \\
    Non-Normal & Cover & Frequency & Gain, Saabas, and SHAP \\
    \hline
    \end{tabular}
\end{table}

\begin{table}
    \caption[Mean and standard error of power from $50$ normal distributed simulation tests]{Mean and standard error of \textbf{power} from $50$ \textbf{normal} distributed simulation tests for ranking test statistics and knockoff types, where $n=100$, $p=500$, $\pi=0.04$, signal strength $=1.5$, design matrix variance $\sigma^2=1$, and $q=1000$.}
    \centering
    \begin{tabular}{c|cccc}
    \midrule\midrule
     & Sparse Gaussian & Shrunk Gaussian & $10$-Principal Component & $30$-Principal Component \\
    \midrule\midrule
    Cover & $0.042$ $(0.014)$ & $0.460$ $(0.018)$ & $0.448$ $(0.018)$ & $0.297$ $(0.026)$ \\
    Frequency & $0.054$ $(0.016)$ & $0.480$ $(0.017)$ & $0.463$ $(0.017)$ & $0.355$ $(0.021)$ \\
    Gain & $0.007$ $(0.007)$ & $0.423$ $(0.016)$ & $0.398$ $(0.017)$ & $0.211$ $(0.021)$ \\
    Saabas & $0.037$ $(0.013)$ & $0.471$ $(0.017)$ & $0.448$ $(0.017)$ & $0.283$ $(0.024)$ \\
    SHAP & $0.037$ $(0.013)$ & $0.467$ $(0.017)$ & $0.448$ $(0.017)$ & $0.284$ $(0.024)$ \\
    \midrule\midrule
    \end{tabular}
    \label{kobt_norm_power}
\end{table}

\begin{table}
    \caption[Mean and standard error of FDR from $50$ normal distributed simulation tests]{Mean and standard error of \textbf{FDR} from $50$ \textbf{normal} distributed simulation tests for ranking test statistics and knockoff types, where $n=100$, $p=500$, $\pi=0.04$, signal strength $=1.5$, design matrix variance $\sigma^2=1$, and $q=1000$.}
    \centering
    \begin{tabular}{c|cccc}
    \midrule\midrule
     & Sparse Gaussian & Shrunk Gaussian & $10$-Principal Component & $30$-Principal Component \\
    \midrule\midrule
    Cover & $0.091$ $(0.031)$ & $0.721$ $(0.008)$ & $0.693$ $(0.011)$ & $0.481$ $(0.035)$ \\
    Frequency & $0.139$ $(0.038)$ & $0.751$ $(0.008)$ & $0.739$ $(0.009)$ & $0.640$ $(0.023)$ \\
    Gain & $0.007$ $(0.007)$ & $0.684$ $(0.010)$ & $0.660$ $(0.012)$ & $0.407$ $(0.039)$ \\
    Saabas & $0.070$ $(0.026)$ & $0.720$ $(0.008)$ & $0.694$ $(0.011)$ & $0.450$ $(0.035)$  \\
    SHAP & $0.070$ $(0.026)$ & $0.719$ $(0.009)$ & $0.694$ $(0.011)$ & $0.448$ $(0.035)$ \\
    \midrule\midrule
    \end{tabular}
    \label{kobt_norm_fdr}
\end{table}

\begin{table}
    \caption[Mean and standard error of power from $50$ Poisson-distributed simulation tests]{Mean and standard error of \textbf{power} from $50$ \textbf{Poisson}-distributed simulation tests for ranking test statistics and knockoff types, where $n=100$, $p=500$, $\pi=0.04$, signal strength $=1.5$, design matrix variance $\sigma^2=1$, and $q=1000$.}
    \centering
    \begin{tabular}{c|cccc}
    \midrule\midrule
     & Sparse Gaussian & Shrunk Gaussian & $10$-Principal Component & $30$-Principal Component \\
    \midrule\midrule
    Cover & $0.010$ $(0.010)$ & $0.380$ $(0.015)$ & $0.359$ $(0.016)$ & $0.105$ $(0.023)$ \\
    Frequency & $0.008$ $(0.008)$ & $0.377$ $(0.015)$ & $0.338$ $(0.017)$ & $0.086$ $(0.021)$ \\
    Gain & $0.008$ $(0.008)$ & $0.332$ $(0.015)$ & $0.262$ $(0.023)$ & $0.053$ $(0.017)$ \\
    Saabas & $0.008$ $(0.008)$ & $0.349$ $(0.015)$ & $0.276$ $(0.023)$ & $0.040$ $(0.015)$ \\
    SHAP & $0.008$ $(0.008)$ & $0.351$ $(0.015)$ & $0.275$ $(0.023)$ & $0.041$ $(0.015)$ \\
    \midrule\midrule
    \end{tabular}
    \label{kobt_pois_power}
\end{table}

\begin{table}
    \caption[Mean and standard error of FDR from $50$ Poisson-distributed simulation tests]{Mean and standard error of \textbf{FDR} from $50$ \textbf{Poisson}-distributed simulation tests for ranking test statistics and knockoff types, where $n=100$, $p=500$, $\pi=0.04$, signal strength $=1.5$, design matrix variance $\sigma^2=1$, and $q=1000$.}
    \centering
    \begin{tabular}{c|cccc}
    \midrule\midrule
     & Sparse Gaussian & Shrunk Gaussian & $10$-Principal Component & $30$-Principal Component \\
    \midrule\midrule
    Cover & $0.007$ $(0.007)$ & $0.551$ $(0.016)$ & $0.488$ $(0.021)$ & $0.142$ $(0.031)$ \\
    Frequency & $0.007$ $(0.007)$ & $0.545$ $(0.015)$ & $0.474$ $(0.024)$ & $0.116$ $(0.029)$ \\
    Gain & $0.005$ $(0.005)$ & $0.496$ $(0.022)$ & $0.396$ $(0.034)$ & $0.084$ $(0.027)$ \\
    Saabas & $0.005$ $(0.005)$ & $0.514$ $(0.019)$ & $0.388$ $(0.032)$ & $0.066$ $(0.024)$ \\
    SHAP & $0.005$ $(0.005)$ & $0.514$ $(0.019)$ & $0.392$ $(0.032)$ & $0.065$ $(0.024)$ \\
    \midrule\midrule
    \end{tabular}
    \label{kobt_pois_fdr}
\end{table}

\section{Real Data Applications}
\label{sec:app}

\subsection{Tumor Sample Purity Estimation} \label{purity_est}

Tumor sample purity refers to the percentage of cancer cells in a tumor tissue sample. It is used to discover the roles of cancerous and non-cancerous cells in the tumour microenvironment, which mainly comprises immune cells \citep{aran2015systematic, turley2015immunological}. To estimate tumor purity, \cite{carter2012absolute} proposed ABSOLUTE, a method used to perform analysis of somatic DNA alterations. In addition, it has been reported that DNA methylation data and expression data from selected stromal genes \citep{houseman2012dna, yoshihara2013inferring} have been successfully used for estimation. Recently, \cite{aran2015systematic} and \cite{li2019putative} used RNA-seq gene expression data to determine tumor purity and found several gene signatures of individual cancer types. As it has been shown reasonable to estimate tumor sample purity using gene expression data, we apply our KOBT algorithm so that we can compare our results with those of previous reports. Here, our interest is not the accuracy of estimation but the selection of genes that are important in estimation with no topology assumptions. If a gene's expression level is positively correlated with tumor purity, then it is highly likely that this gene is expressed primarily by cancer cells in tumor samples.

Processed TCGA RNA-seq gene expression data was downloaded from the Pan-Cancer Atlas Publication website \citep{hoadley2018cell}. Among all $33$ available tumor types, breast invasive carcinoma (BRCA) and skin cutaneous melanoma (SKCM) were chosen for tumor purity estimation. There are $1,017$ BRCA and $459$ SKCM samples with $17,176$ expressed genes for each sample. The genes with missing values or zero variance were filtered out. We used the tumor purity estimates in \cite{hoadley2018cell} as the responses, which were obtained using ABSOLUTE. The responses are thus bounded between $0$ and $1$. 

To reduce the original dimensionality to a lower value, we apply Lasso \citep{tibshirani1996regression} on original data sets before performing our KOBT steps. After tuning the penalty terms in Lasso objective functions, $486$ and $496$ genes of BRCA and SKCM respectively are selected for further analysis. We generated $q=300$ knockoffs for each gene. Here, we compare two knockoff types: shrunk Gaussian and $10$-principal component knockoffs. We conducted Bayesian optimization and $10$-fold cross-validation to choose the optimal parameters for the boosted tree algorithm (see Section \ref{sec:para} for details of the parameters). We used the SHAP for importance evaluation. Therefore, our test statistic is the average difference of SHAP values between each gene and its own knockoff, as shown in Equation \ref{sead}. Finally, FDR control is applied to these test statistics of genes with a targeted FDR level of $10\%$.

\begin{table}
    \caption[Genes whose expression is related to BRCA tumor purity]{Genes whose expression is related to BRCA tumor purity, where targeted FDR $=0.1$. Genes are selected using the Shrunk Gaussian knockoff. Overlapping genes with ESTIMATE models are in \textbf{bold}.}
    \centering
    \begin{tabular}{l}
    \midrule\midrule
    Detected Genes \\
    \hline
    C1S, C2orf48, \textbf{CCDC69}, CSF2RB, CXorf65, DNAJC12, EDN3, FAM163A, FAM65B, \\
    \textbf{FGR}, GBP3, HGFAC, HTR2A, \textbf{IL7R}, KIAA0087, NCRNA00175, NRADDP, NRG2, \\
    PM20D1, PNOC, SLAIN1, UNQ6494 \\
    \midrule\midrule
    \end{tabular}
    \label{brca_purity}
\end{table}

\begin{figure}[!htbp]
  \centering
  \includegraphics[scale=0.7]{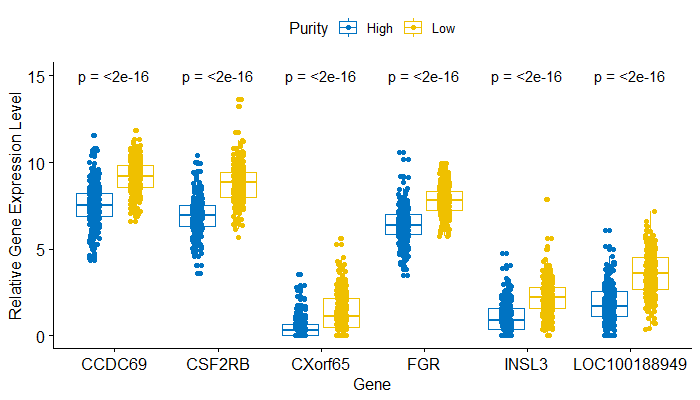}
  \caption[Relative gene expression level comparison for selected genes in low and high purity samples of BRCA]{Relative gene expression level comparison for selected genes in low (yellow) and high (blue) purity samples of BRCA. The non-parametric Wilcoxon signed-rank test is conducted for each low-high pair. Their corresponding \textit{p}-values are included at the top of the plot. The relative gene expression level is the $\log_2$-transformed normalized expression read counts mapped (unit: million reads).}
  \label{fig:brca_highlow}
\end{figure}

\begin{figure}[!htbp]
  \centering
  \includegraphics[scale=0.4]{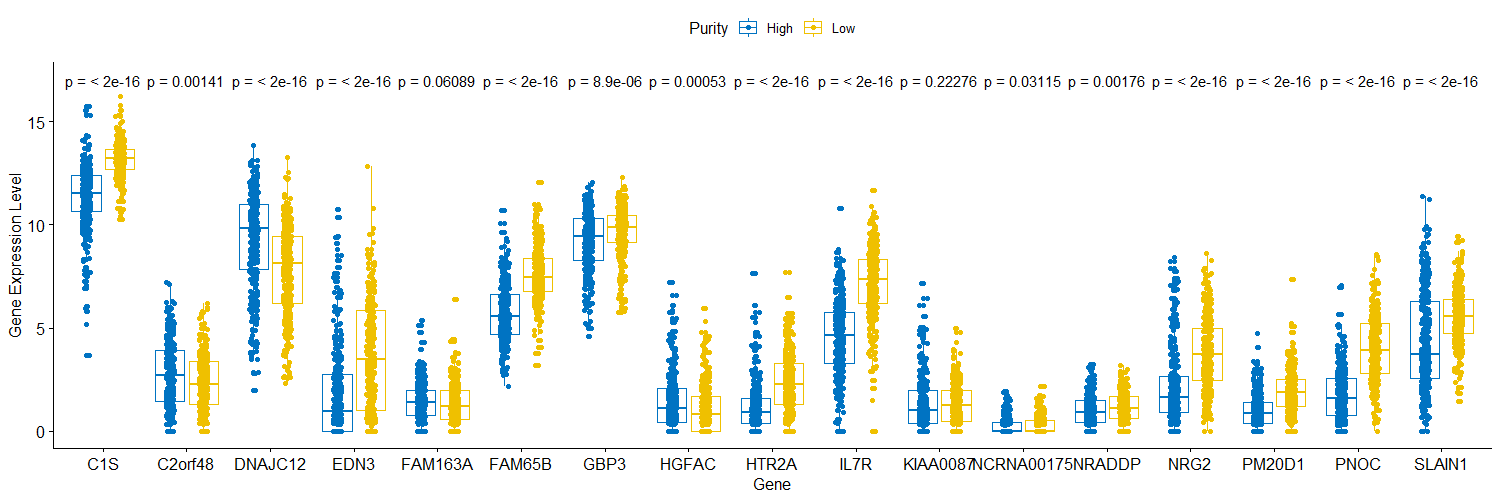}
  \caption[Relative gene expression level comparison for other genes in low and high purity samples of BRCA]{Relative gene expression level comparison for other genes in low (yellow) and high (blue) purity samples of BRCA. The  non-parametric Wilcoxon signed-rank test is conducted for each low-high pair. Their corresponding \textit{p}-values are included at the top of the plot. The relative gene expression level is the $\log_2$-transformed normalized expression read counts mapped (unit: million reads).}
  \label{fig:brca_purity_2}
\end{figure}

\begin{figure}[!htbp]
  \centering
  \includegraphics[scale=0.5]{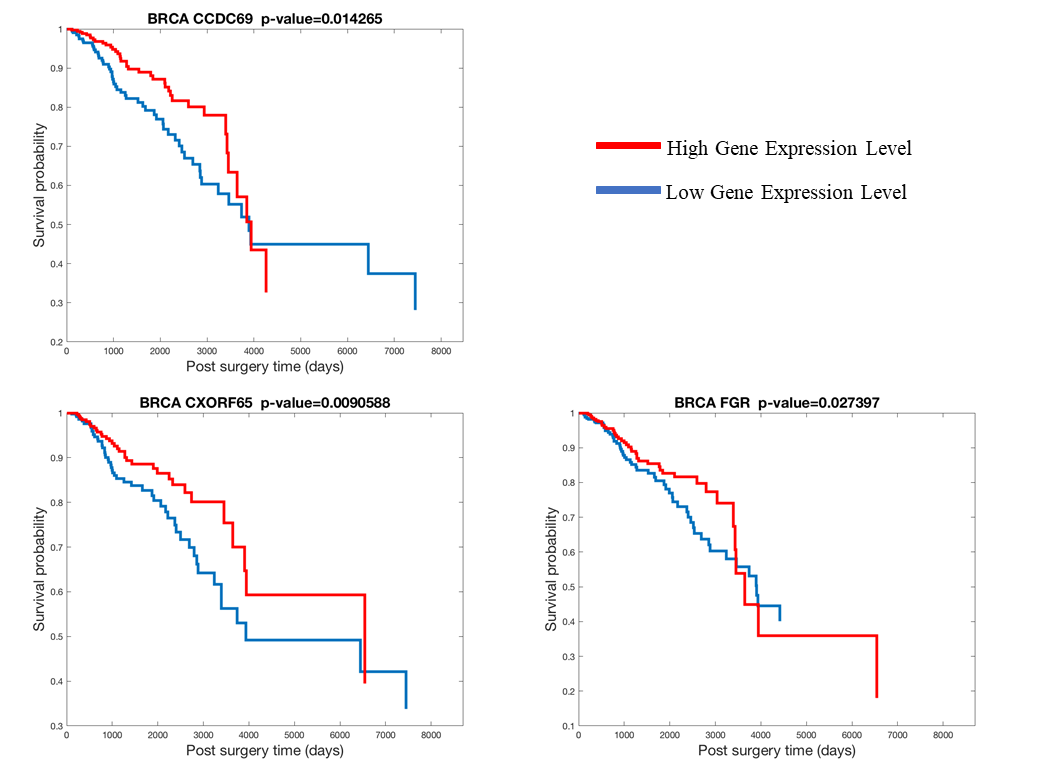}
  \caption[Plots of the Kaplan–Meier estimator]{Plots of the Kaplan–Meier estimators and \textit{p}-values for coefficients in Cox regression of BRCA. Samples with high gene expression levels are in red, and samples with low gene expression levels are in blue.}
  \label{fig:brca_survival}
\end{figure}

Table \ref{brca_purity} reports the selected genes whose expression is related to BRCA tumor purity, where the targeted FDR is $0.1$. Genes are selected using the Shrunk Gaussian knockoff. Genes selected by the $10$-principal component knockoff are listed in Table \ref{10PC}. Among all detected genes, CSF2RB (colony stimulating factor 2 receptor beta common subunit), C1S (complement C1s), CCDC69 (coiled-coil domain containing 69), and FGR are also reported among the top $10$-ranked important genes for pan-cancer tumor purity prediction in \cite{li2019putative}. CSF2RB is an immune-related gene. It has been reported that in BRCA, the high expression of CSF2RB is positively correlated with patient survival \citep{liu2019bioinformatic}. ESTIMATE \citep{yoshihara2013inferring} uses stromal and immune gene expression to predict tumor purity. Our detected immune genes CCDC69, FGR, and IL7R are included in \cite{yoshihara2013inferring}. The box plots of gene expression levels for selected genes are presented in Figure \ref{fig:brca_highlow}. These genes are selected because they are detected by both types of knockoffs. Other genes are plotted in Figure \ref{fig:brca_purity_2}. All samples are grouped according to their purity: samples with the top $1/3$ purity are labeled as high (blue); and samples with the bottom $1/3$ purity are labeled as low (yellow). The non-parametric Wilcoxon signed-rank test is conducted for each low-high pair. Their corresponding \textit{p}-values are included at the top of the plot. It is shown that genes expression levels of almost all selected genes are highly correlated with tumor purity. Samples are grouped according to their gene expression levels. The top $1/3$ samples are grouped into the high-level group, and the bottom $1/3$ samples are grouped into the low-level group. Survival analysis has been conducted for these genes. Among them, CCDC69, CXORF65, and FGR have significant \textit{p}-values for coefficients in Cox regression. Plots of the Kaplan–Meier estimators and \textit{p}-values are in Figure \ref{fig:brca_survival}.

\begin{figure}[!htbp]
  \centering
  \includegraphics[scale=0.7]{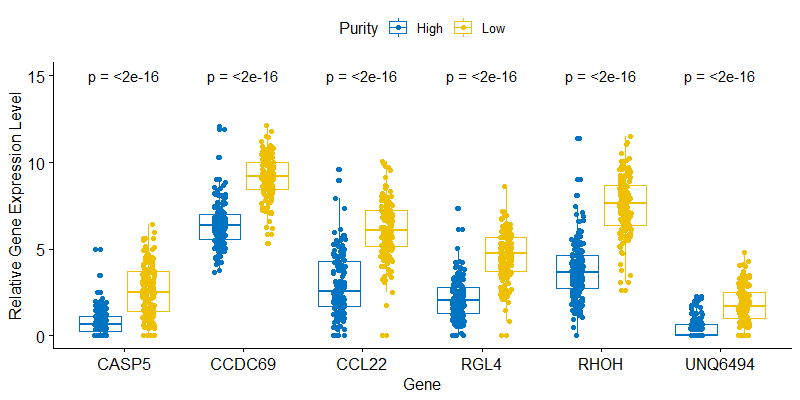}
  \caption[Relative gene expression level comparison for selected genes in low and high purity samples of SKCM]{Relative gene expression level comparison for selected genes in low (yellow) and high (blue) purity samples of SKCM. The  non-parametric Wilcoxon signed-rank test is conducted for each low-high pair. Their corresponding \textit{p}-values are included at the top of the plot. The relative gene expression level is the $\log_2$-transformed normalized expression read counts mapped (unit: million reads).}
  \label{fig:skcm_highlow}
\end{figure}

\begin{figure}[!htbp]
  \centering
  \includegraphics[scale=0.4]{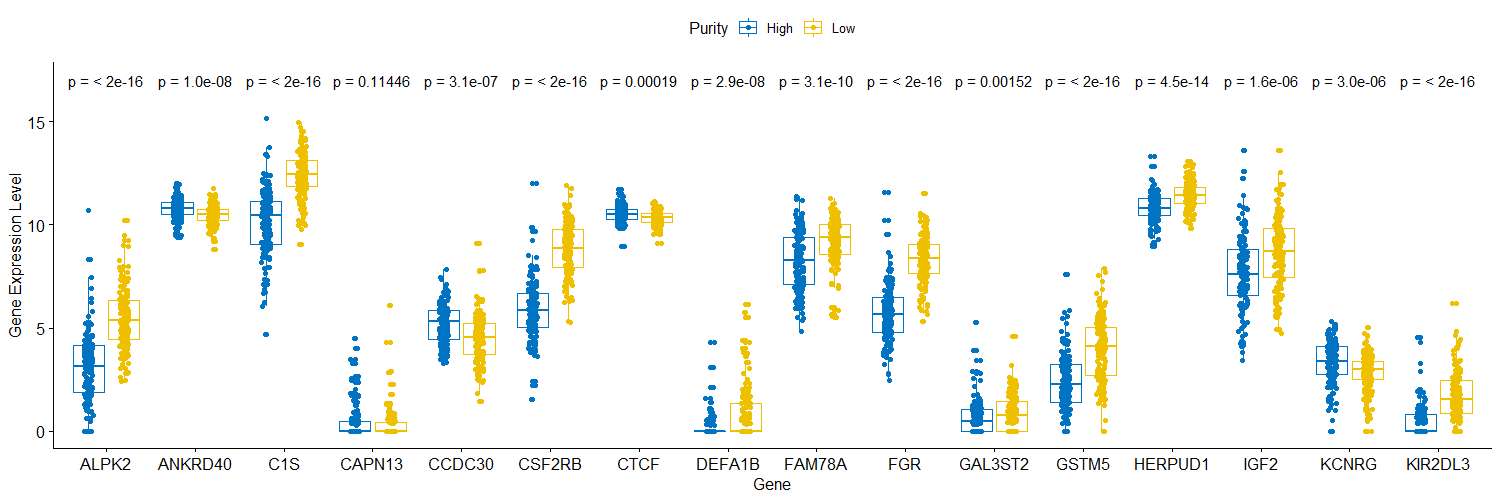}
  \includegraphics[scale=0.4]{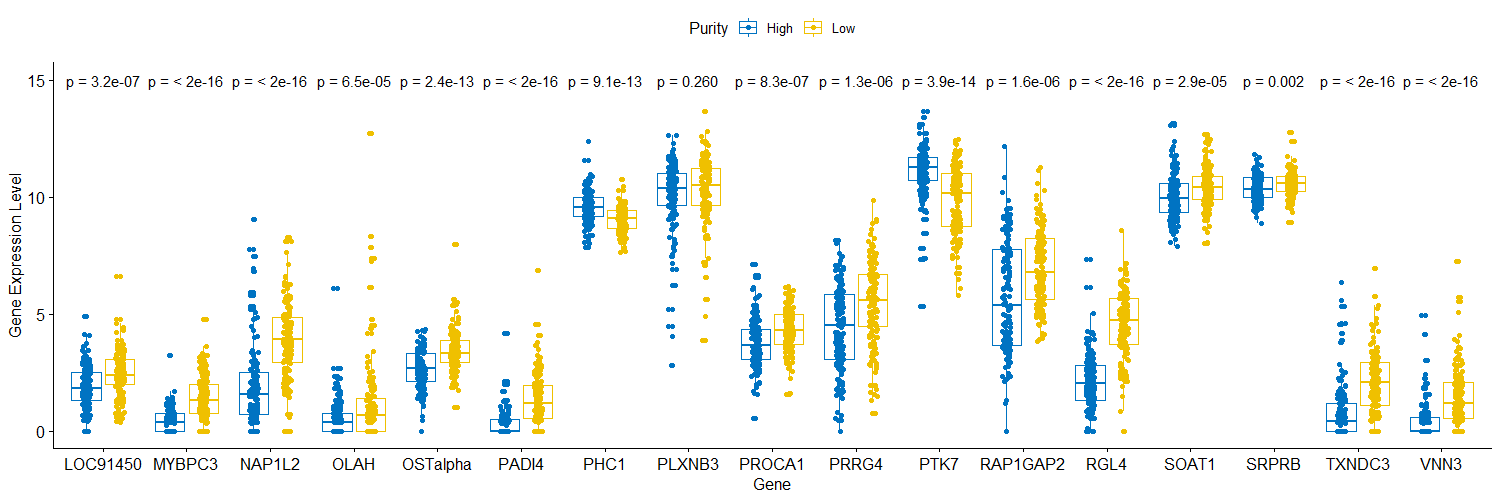}
  \caption[Relative gene expression level comparison for other genes in low and high purity samples of SKCM]{Relative gene expression level comparison for other genes in low (yellow) and high (blue) purity samples of SKCM. The non-parametric Wilcoxon signed-rank test is conducted for each low-high pair. Their corresponding \textit{p}-values are included at the top of the plot. The relative gene expression level is the $\log_2$-transformed normalized expression read counts mapped (unit: million reads).}
  \label{fig:skcm_purity_2}
\end{figure}

Table \ref{skcm_purity} reports more genes associated with SKCM tumor purity. \cite{li2019putative} reports CSF2RB, RHOH, C1S, CCDC69, and CCL22 as among the top 10-ranked important genes for pan-cancer tumor purity prediction. Most of these detected genes are from stromal cells and not cancer cells. For example, the expression level of CCDC69 is negatively correlated with tumor purity, which can not be true if it mainly comes from cancer cells. Immune genes CCDC69, FGR, and RHOH, and stromal gene TXNDC3 are included in ESTIMATE models. The box plots of gene expression levels for selected genes are presented in Figure \ref{fig:skcm_highlow}. Other genes are plotted in Figure \ref{fig:skcm_purity_2}. Same as it is for BRCA, samples with the top $1/3$ purity are labeled as high (blue); and samples with the bottom $1/3$ purity are labeled as low (yellow). Related \textit{p}-values from Wilcoxon signed-rank tests are attached to the plots. We see that genes expression levels of almost all selected genes are highly correlated with tumor purity. All samples are grouped according to their gene expression levels. Plots of their Kaplan–Meier estimators and \textit{p}-values are in Figure \ref{fig:skcm_survival_1} and \ref{fig:skcm_survival_2}. In summary, for tumor purity estimation, we propose that our KOBT algorithm can detect a few genes that are largely expressed in stromal cells. Our results reproduce previously reported work using different algorithms.

\begin{table}
    \caption[Genes whose expression is related to SKCM tumor purity]{Genes whose expression is related to SKCM tumor purity, where targeted FDR $=0.1$. Genes are selected using the Shrunk Gaussian knockoff. Overlapping genes with ESTIMATE models are in \textbf{bold}.}
    \centering
    \begin{tabular}{l}
    \midrule\midrule
    Detected Genes \\
    \hline
    CSF2RB, \textbf{RHOH}, C1S, \textbf{CCDC69}, CCL22, \textbf{FGR}, ALPK2, ANKRD40, CAPN13, CASP5, \\
    CCDC30, CTCF, DEFA1B, FAM78A, GAL3ST2, GSTM5, HERPUD1, IGF2, KCNRG, \\
    KIR2DL3, LOC100129066, LOC91450, MYBPC3, NAP1L2, OLAH, OSTalpha, PADI4, \\
    PHC1, PLXNB3, PROCA1, PRRG4, PTK7, RAP1GAP, RGL4, SOAT1, SRPRB, \\
    \textbf{TXNDC3}, VNN3 \\
    \midrule\midrule
    \end{tabular}
    \label{skcm_purity}
\end{table}

\subsection{Tumor Type Classification}

For application of KOBT on classification, we focus on its performance of assigning tumors to known classes and identifying those genes whose expression levels are related to the classification. In 1999, \cite{golub1999molecular} used gene expression monitoring by DNA microarrays to conduct cancer classification. Their test on classifying acute myeloid leukemia (AML) and acute lymphoblastic leukemia (ALL) showed that it was realistic to distinguish different cancer subtypes only based on gene expression data. \cite{li2017comprehensive} performed a pan-cancer classification with RNA-seq data from TCGA using boosted tree and $k$-nearest neighbours methods. Since KOBT algorithm can detect signals with false discovery rate control, we would like to see how it works on gene selection for tumor type classification.

The TCGA RNA-seq gene expression data is the same as in Subsection \ref{purity_est}. Instead of a pan-cancer classification, we focus on two challenging binary classifications for (1) esophageal carcinoma (ESCA) \textit{vs} stomach adenocarcinoma (STAD) and (2) rectum adenocarcinoma (READ) \textit{vs} colon adenocarcinoma (COAD). We choose these two pairs of cancers because of their similarities. Esophageal carcinoma (ESCA) and stomach adenocarcinoma (STAD) are both malignant tumors in the digestive tract. In \cite{li2017comprehensive}, it has been reported that almost all READ samples were mis-assigned to COAD. We have $499$ samples for the ESCA vs STAD classification test and $529$ samples for the READ vs COAD one. The steps are the same as in Subsection \ref{purity_est} except that the responses are binary now: $0$ for one cancer and $1$ for another cancer.

\begin{table}
    \caption[Genes whose expression is related to ESCA vs STAD tumor classification]{Genes whose expression is related to ESCA vs STAD tumor classification, where targeted FDR $=0.1$.}
    \centering
    \begin{tabular}{|l|}
    \hline
    Detected Genes \\
    \hline
    BARX1, HAND2, KRT14, NVL \\
    \hline
    \end{tabular}
    \label{evs_vs}
\end{table}

\begin{figure}[!htbp]
  \centering
  \includegraphics[scale=0.6]{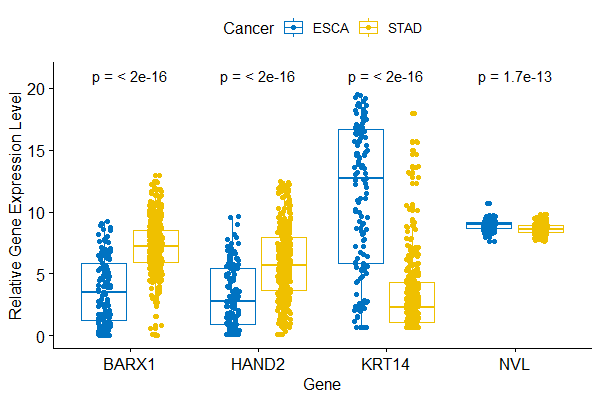}
  \caption[Relative gene expression level comparison for selected genes in ESCA and STAD samples]{Relative gene expression level comparison for selected genes in ESCA (blue) and STAD (yellow) samples. The non-parametric Wilcoxon signed-rank test is conducted. The corresponding \textit{p}-values are included at the top of the plot. The relative gene expression level is the $\log_2$-transformed normalized expression read counts mapped (unit: million reads).}
  \label{fig:evs_plot}
\end{figure}

In Table \ref{evs_vs}, the selected genes that can distinguish ESCA and STAD are presented. Among them, BARX1 is a stomach mesenchymal transcription factor. \cite{kim2005stomach} has shown that BARX1 loss in the mesenchyme prevents stomach epithelial differentiation of overlying endoderm and induces intestine-specific genes instead. Their results defined a transcriptional and signaling pathway of inductive cell interactions in vertebrate organogenesis. \cite{kim2011regulation} proved that BARX1 controls mouse stomach morphogenesis and is required to specify stomach-specific epithelium in adjacent endoderm. HAND2 is an RNA Gene, which is affiliated with the long non-coding RNA class. \cite{tsubokawa2002heterogeneity} stated that KRT14 is often expressed by tumor cells in the trabecular nests of the primary carcinoma. The box plots of gene expression levels for all detected genes are presented in Figure \ref{fig:evs_plot}. All samples are grouped according to their cancer types: ESCA samples are in blue, and STAD samples are in yellow. The non-parametric Wilcoxon signed-rank test is conducted for each gene. Their corresponding \textit{p}-values are included at the top of the plot. It is shown that genes expression levels of all selected genes are highly correlated with the cancer type. 

\begin{figure}[!htbp]
  \centering
  \includegraphics[scale=0.6]{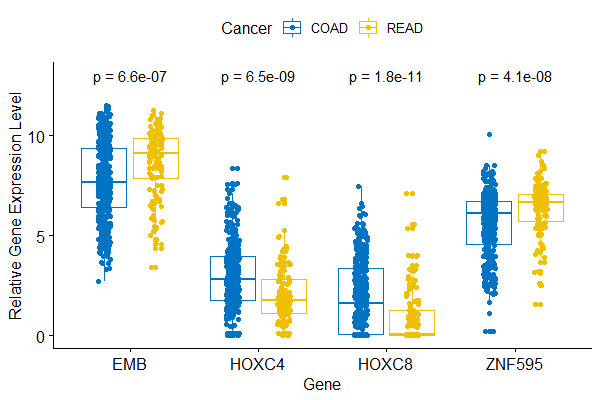}
  \caption[Relative gene expression level comparison for selected genes in COAD and READ samples]{Relative gene expression level comparison for selected genes in COAD (blue) and READ (yellow) samples. The non-parametric Wilcoxon signed-rank test is conducted. The corresponding \textit{p}-values are included at the top of the plot. The relative gene expression level is the $\log_2$-transformed normalized expression read counts mapped (unit: million reads).}
  \label{fig:rvc_plot}
\end{figure}

In Table \ref{rvc_vs}, we show the detected genes that can used for READ and COAD classification. HOXC4 and HOXC8 are members of Homeobox genes, which is a large family of transcription factors that direct the formation of many body structures during early embryonic development. It has been observed that HOXC family gene expression is upregulated in most solid tumor types \citep{bhatlekar2014hox}. The box plots of gene expression levels for all detected genes are presented in Figure \ref{fig:rvc_plot}. All samples are grouped according to their cancer types: COAD samples are in blue, and READ samples are in yellow. The non-parametric Wilcoxon signed-rank test is conducted for each gene. Their corresponding \textit{p}-values are included at the top of the plot. It is shown that genes expression levels of all selected genes are highly correlated with the cancer type. 

\begin{table}
    \caption[Genes whose expression is related to READ vs COAD tumor classification]{Genes whose expression is related to READ vs COAD tumor classification, where targeted FDR $=0.1$.}
    \centering
    \begin{tabular}{|l|}
    \hline
    Detected Genes \\
    \hline
    EMB, HOXC4, HOXC8, ZNF595 \\
    \hline
    \end{tabular}
    \label{rvc_vs}
\end{table}

Unlike in tumor purity estimation, cancer genes play a more important role in tumor type classification. We detect some protein coding genes and find previous published literature to support our findings in both classification tests. Given the fact that the mechanism of cancers is complicated, and the cancer types we choose have been reported to be hard to distinguish, we show that our proposed KOBT algorithm to be robust in real data with unknown models.

\section{Conclusions}
\label{sec:con}

In this article, we introduce a new model-free variable selection method, called knockoff boosted tree, KOBT. Given the nature of boosted tree models, no prior model topology knowledge is required. We extend the current application of knockoff methods to tree models. Meanwhile, we proposed two new sampling methods to generate knockoffs, the principal component construction knockoff and the sparse Gaussian knockoff. Unlike currently available methods, the principal component construction knockoff does not depend on Gaussian assumption of design matrix. To evaluate the power of our generated knockoffs in simulation tests, we define a new test statistic, \textit{mean absolute angle of columns}, which stands for the average distance between column vectors in two matrices. We apply \textit{kernel maximum mean discrepancy} to test whether our proposed knockoffs are drawn from the same distributions as the original vectors. In our boosted tree framework, we fix most hyperparameters at multiple reasonable levels and leave regularization parameters to be tuned by Bayesian optimization. We consider different importance scores in tree models. Combinations of different knockoffs and importance test statistics are listed in our results. We test our strategy on different unknown true models, including main effect, interaction, exponential, and second order models. Finally, we apply our algorithm on tumor purity estimation and tumor type classification using TCGA gene expression data. We see that KOBT performs well and our results match previously published literature. Finally, we provide an R package to implement our proposed approaches, which is available at \url{https://cran.r-project.org/web/packages/KOBT/index.html}.

\section*{Acknowledgements}
This research was supported by the Intramural Research Program of the NIH, National Institute of Environmental Health Sciences.

\clearpage
\section{Supplementary Information}

\subsection{Parameters in Real Data Applications}
\label{sec:para}

\begin{table}[h]
    \caption[XGBoost parameters in real data applications]{XGBoost parameters in real data applications.}
    \centering
    \begin{tabular}{l|l}
    \midrule\midrule
    Parameter & Value \\
    \midrule\midrule
    Booster & gbtree \\
    \hline
    Maximum number of trees & $5000$ \\
    \hline
    Learning rate & $0.01$ \\
    \hline
    Maximum tree depth & $6$ \\
    \hline
    Minimum leaf weight & $10$ \\
    \hline
    Sub-sample & $80\%$ \\
    \hline
    Sub-feature & $80\%$ \\
    \hline
    $\gamma$ & $[0,20]$ (a tuning region for Bayesian optimization) \\
    \hline
    $\lambda$ & $[0,20]$ (a tuning region for Bayesian optimization) \\
    \hline
    $\alpha$ & $[0,20]$ (a tuning region for Bayesian optimization) \\
    \hline
    Objective function & reg:squarederror \& binary:logistic \\
    \hline
    Evaluation metric & root mean square error \& classification error \\
    \midrule\midrule
    \end{tabular}
\end{table}

The early stopping rule is stopping adding new trees when average test loss in cross validation does not improve in five additional trees.

\subsection{Selected Genes Using 10-principal Component Knockoff}

\begin{table}[h]
    \caption[Genes whose expression is related to tumor purity using $10$-principal component knockoff]{Genes whose expression is related to tumor purity using $10$-principal component knockoff, where targeted FDR $=0.1$.}
    \centering
    \begin{tabular}{c|l}
    \midrule\midrule
    Cancer & Detected Genes \\
    \midrule\midrule
    BRCA &  CSF2RB,  CCDC69,  FGR,  CXorf65, INSL3,  UNQ6494 \\
    \hline
    \multirow{10}{*}{SKCM} &  RHOH,  CCDC69,  CCL22,  ALPK2,  ANKRD40, BBS1, BEST4, \\
    & C11orf70, C2orf27A,  CAPN13,  CASP5,  CCDC30, CHDH, CHRM4, \\
    & CTAGE9,  CTCF, CXCL6,  DEFA1B, EIF1B, FAHD1,  FAM78A, FBXW8, \\
    & FERMT1,  GAL3ST2, GRM4,  GSTM5,  IGF2,  KCNRG, KDM4D, \\
    & KHDC1,  KIR2DL3, KLF11, LAMA1, LOC100129034,  LOC100129066, \\
    & LOC91450, MIA, MYBPC3, NAP1L2,  OLAH,  OSTalpha,  PADI4, \\
    & PCCB, PER1, PER2, PHC1,  PLXNB3,  PROCA1,  PRRG4,  PTK7, \\
    & PTPN5, PTRH2, PYROXD2,  RAP1GAP2,  RGL4, RIC3, RNF126P1, \\
    & RYBP, SEC61A1, SGSM2, SMAD2,  SOAT1,  SRPRB, TXNDC3, \\
    & USP30, VNN3, ZIC1 \\
    \midrule\midrule
    \end{tabular}
    \label{10PC}
\end{table}

\clearpage
\subsection{Relative Gene Expression Level Comparison in Low and High Purity Samples}

\clearpage
\subsection{Survival Analysis for Other Selected Genes in SKCM}

\begin{figure}[!htbp]
  \centering
  \includegraphics[scale=0.5]{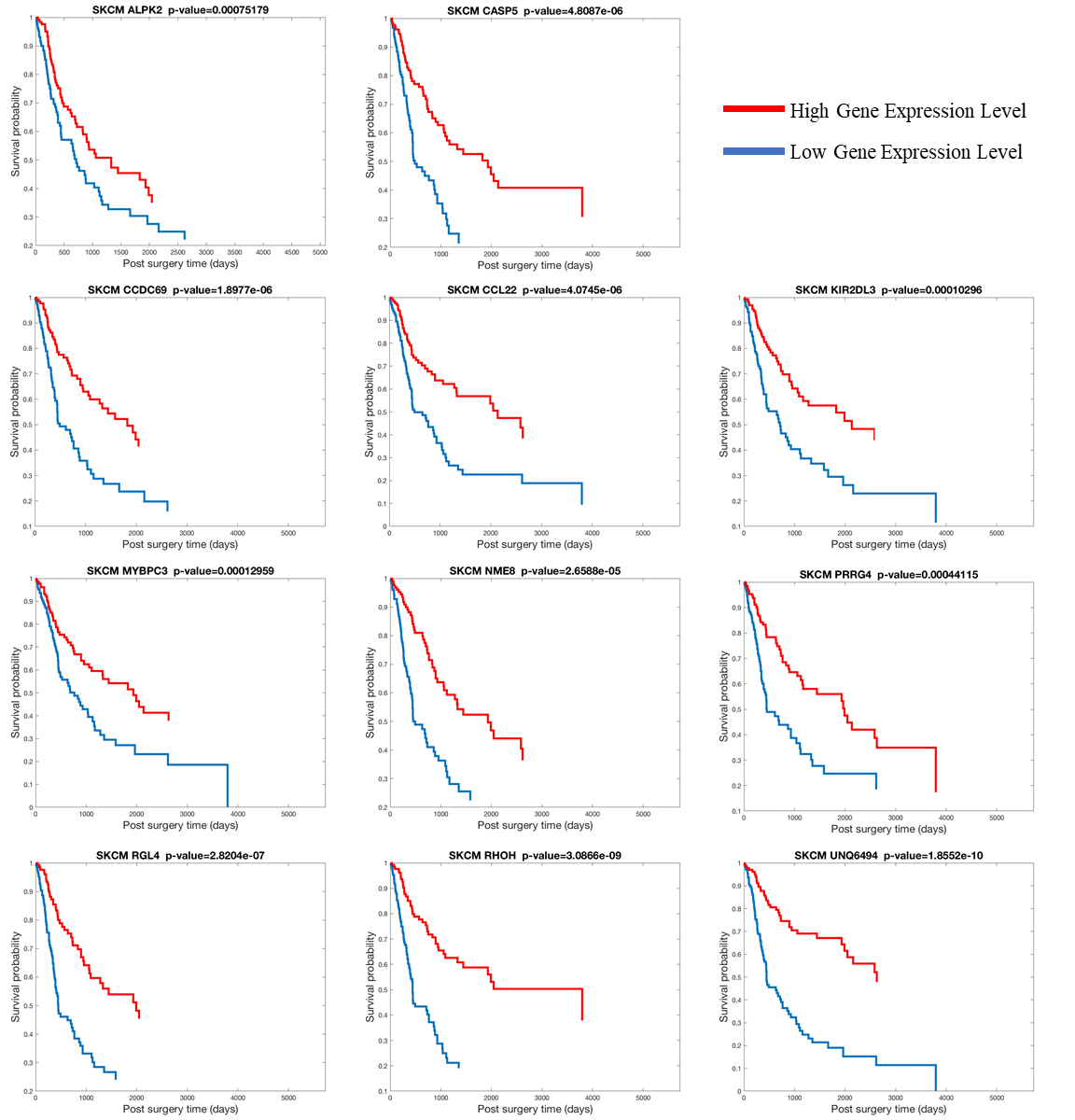}
  \caption[Plots of the Kaplan–Meier estimator]{Plots of the Kaplan–Meier estimators and \textit{p}-values for coefficients in Cox regression of SKCM.}
  \label{fig:skcm_survival_1}
\end{figure}

\begin{figure}[!htbp]
  \centering
  \includegraphics[scale=0.5]{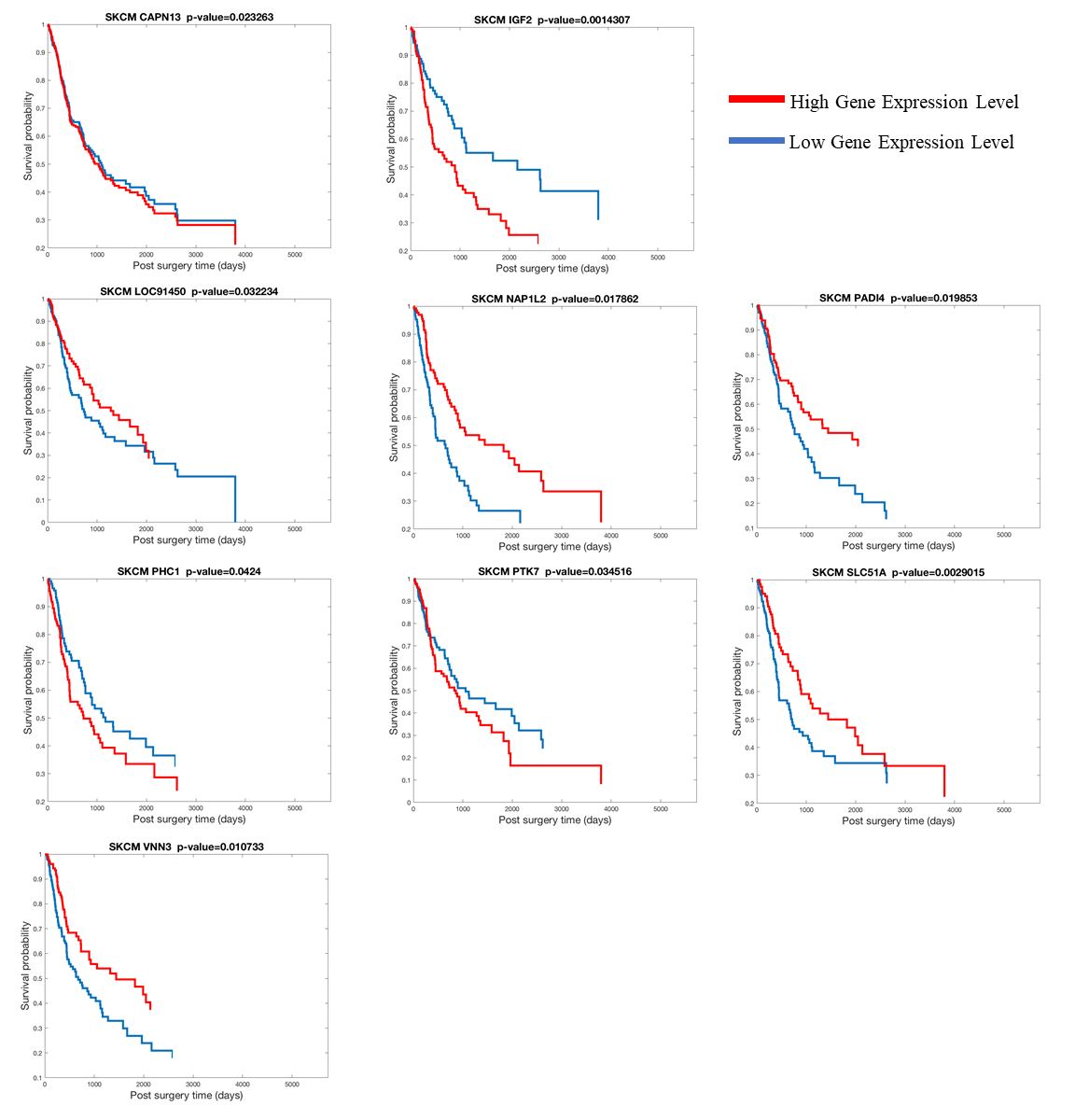}
  \caption[(Continued) Plots of the Kaplan–Meier estimator]{(Continued) Plots of the Kaplan–Meier estimators and \textit{p}-values for coefficients in Cox regression of SKCM.}
  \label{fig:skcm_survival_2}
\end{figure}

\clearpage
\bibliographystyle{apalike}  
\bibliography{references}  


\end{document}